\documentclass[structabstract]{aa}  
\usepackage{graphicx}                   
\usepackage{epstopdf}
\usepackage{amsmath}           

\usepackage{txfonts}
 \usepackage{natbib,twoopt}
 \usepackage[breaklinks=true]{hyperref} 
 \bibpunct{(}{)}{;}{a}{}{,} 
 \newcommandtwoopt{\citeads}[3][][]{\href{http://adsabs.harvard.edu/abs/#3}{\citealp[#1][#2]{#3}}}
 \newcommandtwoopt{\citepads}[3][][]{\href{http://adsabs.harvard.edu/abs/#3}{\citep[#1][#2]{#3}}}
 \newcommandtwoopt{\citetads}[3][][]{\href{http://adsabs.harvard.edu/abs/#3}{\citet[#1][#2]{#3}}} 
 \newcommandtwoopt{\citeyearads}[3][][]{\href{http://adsabs.harvard.edu/abs/#3}{\citeyear[#1][#2]{#3}}} 

\usepackage{lscape}
\newcommand{\chandra}{{\sl Chandra}}
\newcommand{\xmm}{{\sl XMM-Newton}}
\newcommand{\rosat}{{\sl ROSAT}}

\def\hi{H\,{\sc i}}
\def\hii{H\,{\sc ii}}
\newcommand{\halpha}{H$\alpha$}
\newcommand{\oiii}{[\ion{O}{III}]}
\newcommand{\sii}{[\ion{S}{II}]}
\newcommand{\siihalpha}{[\ion{S}{II}]/H$\alpha$}
\newcommand{\deml}{DEM~L299}
\newcommand{\pp}{\mathrm{p}}
\newcommand{\nn}{\mathrm{n}}
\newcommand{\zero}{\mathrm{0}}

\newcommand{\ee}{\mathrm{e}}
\newcommand{\HH}{\mathrm{H}}
\newcommand{\SNR}{\mathrm{SNR}}
\newcommand{\s}{\mathrm{S}}
\newcommand{\xx}{\mathrm{x}}
\newcommand{\bb}{\mathrm{B}}
\newcommand{\TT}{\mathrm{T}}

\newcommand{\LMC}{\mathrm{LMC}}
\newcommand{\thermal}{\mathrm{th}}
\newcommand{\SB}{\mathrm{SB}}
\newcommand{\poly}{\mathrm{poly}}
\newcommand{\tot}{\mathrm{tot}}

\begin{document}
  \title{Multi-frequency study of \object{DEM~L299} in the Large Magellanic Cloud\thanks{Based on observations obtained with XMM-Newton, an ESA science mission
with instruments and contributions directly funded by
ESA Member States and NASA.}}

   \subtitle{}

   \titlerunning{Multi-frequency study of DEM~L299}

   \author{Gabriele Warth\inst{1}
          \and
          Manami Sasaki\inst{1}
          \and
          Patrick J. Kavanagh\inst{1}
          \and
          Miroslav D. Filipovi\'c\inst{2} 
          \and \\
          Sean D. Points\inst{3} 
          \and
          Luke M. Bozzetto\inst{2} 
          }

   \institute{Institut f\"ur Astronomie und Astrophysik, 
              Universit\"at T\"ubingen,
              Sand 1, 
              D-72076 T\"ubingen,
              Germany,
              \email{warth@astro.uni-tuebingen.de}
         \and
              University of Western Sydney,
              Locked Bag 1797,
              Penrith South DC,
              NSW 1979, Australia
         \and
              Cerro Tololo Inter-American Observatory,
              Casilla 603,
              La Serena,
              Chile
             }

   \date{Received ; accepted }

 
   \abstract
   {}
   {We have studied the \ion{H}{II} region \deml\ in the Large
     Magellanic Cloud (LMC) to understand its physical characteristics and morphology in
    different wavelengths.}
 {We performed a spectral analysis of archived \xmm\ EPIC data
  and studied the morphology of \deml\ in X-ray, optical, and radio
  wavelengths. We used H$\alpha$, [\ion{S}{II}], and
  [\ion{O}{III}] data from the Magellanic Cloud Emission Line Survey (MCELS) and
  radio 21~cm line data from the Australia Telescope Compact Array (ATCA) and
  the Parkes telescope as well as radio continuum data (3~cm, 6~cm,
    20~cm, 36~cm) from ATCA and
  from the Molonglo Synthesis
Telescope (MOST).}
 {Our morphological studies imply that, in addition to the supernova remnant SNR
   B0543-68.9 reported in previous studies, a superbubble also
   overlaps the SNR in projection. The position of the SNR is clearly defined through the [\ion{S}{II}]/H$\alpha$ flux
   ratio image. Moreover, the optical images show a shell-like structure that is located
   farther to the north and is filled with diffuse
   X-ray emission, which again indicates the superbubble. Radio
   21~cm line data show a
   shell around both objects. 
   Radio continuum data show diffuse emission at
     the position of \deml, which appears clearly distinguished from the
     \hii\ region LHA~120-N~164 that lies south-west of it. We determined the
     spectral index of SNR B0543-68.9 to be $\alpha=-0.34$, which indicates the
     dominance of thermal emission and therefore a rather mature remnant. We determined the basic properties of the
   diffuse X-ray emission for the SNR, the superbubble, and a possible blowout
   region of the bubble, as suggested by the optical and X-ray data. We
   obtained an age of 
$(8.9 ^{+9.2}_{-5.4})$~kyr 
for the SNR and a
     temperature of $(0.64 ^{+0.73}_{-0.20})$~keV for the hot gas inside the
     SNR, as well as a thermal energy content and temperature of the hot gas
     inside the superbubble of $(4.3 ^{+8.1}_{-2.6})\times
     10^{50}\,\phi^{0.5}$~erg 
and $(0.74^{+0.36}_{-0.30})$~keV, 
with $\phi$ being the gas-filling factor.
 } 
 {
We conclude that \deml\ consists
of a superposition of SNR B0543-68.9 and a superbubble,
which we identified based on optical data. 
   }


   \maketitle
%


\begin{figure*}
      \includegraphics[width=\textwidth]{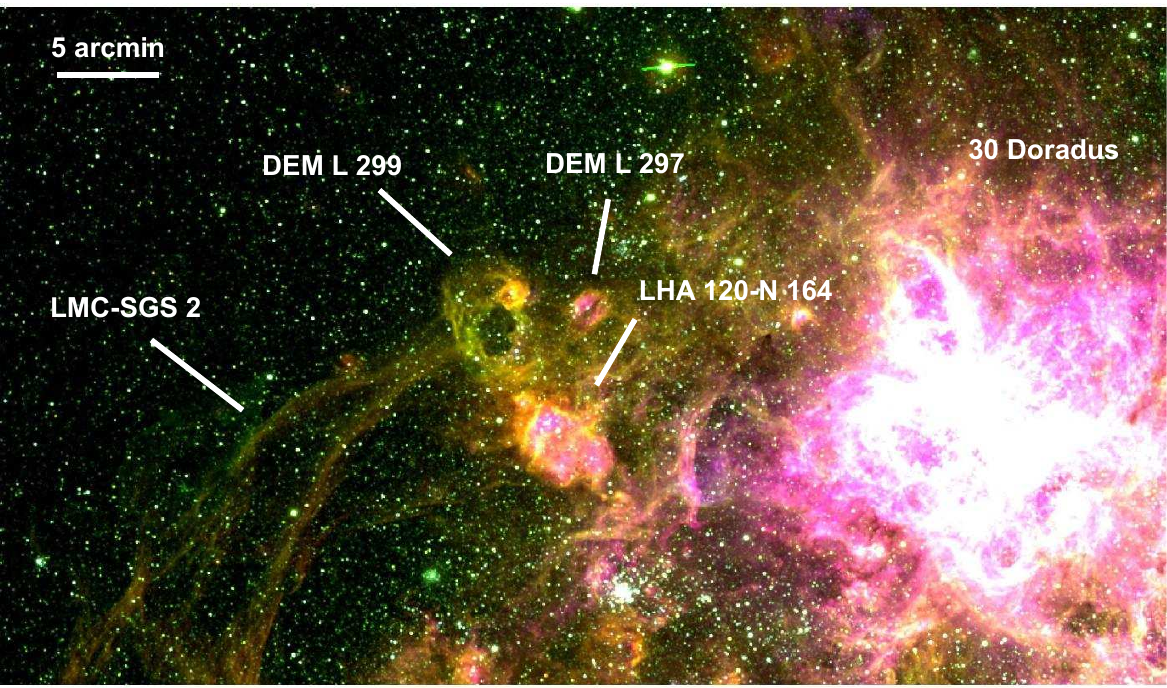}
  \caption{Regions around \deml\ in the LMC with \deml\ visible as a cavity, lying
      north-east of emission region LHA 120-N 164 and east of \object{DEM~L297}. Parts of
      the supergiant shell LMC-SGS~2 can be seen as filaments in the eastern
      part of the image, while 30~Doradus is visible as a bright emission
      region in the west. The image is
      created out of MCELS data and is a three-colour image showing the
      H$\alpha$ emission in red, [\ion{S}{II}] emission in green, and
      [\ion{O}{III}] emission in blue. The whole image shows an area of
      $60\arcmin\,\times\,34\arcmin$ ($\sim$\,870~$\rm pc\,\times\,500~\rm
        pc$). North is approximately in the upward direction, east is to the left.
              }
         \label{FigHalphaBig}
   \end{figure*}

\section{Introduction}
\label{sect_Intro}

The interstellar medium (ISM) can be found mainly in three different phases
\citepads{1977ApJ...218..148M}:
a hot phase with temperatures of $\sim$\,10$^6$~K, created
through stellar winds and supernovae (SNe), observable in
X-rays, for example, as
supernova remnants (SNRs) or the interiors of bubbles, superbubbles, or supergiant shells; a warm phase
with $\sim$\,10$^4$~K, heated by the radiation of hot stars, for instance, observable as \ion{H}{II} regions in optical emission lines; and a cold phase with
temperatures $<$~100~K, which is found for example in molecular clouds or \ion{H}{I} regions,
that are visible through the 21~cm hydrogen line. The ISM plays an important
role for the understanding of the matter cycle, star formation, and stellar and
galactic evolution. 

\deml\ is a complex \ion{H}{II} region located in the Large
Magellanic Cloud (LMC). It is found east of \object{30~Doradus} and towards
the northern rim of the supergiant shell \object{LMC-SGS 2} and is therefore
located close to active star-forming regions (see Fig.~\ref{FigHalphaBig}). The region was catalogued as
an H$\alpha$ emission nebula (N~165) by
\citetads{1956ApJS....2..315H} 
and harbours the supernova remnant SNR B0543-68.9. 
The remnant was classified as an SNR candidate (\deml) in the H$\alpha$ catalogue of
\citetads{1976MmRAS..81...89D} 
and was optically confirmed to be an SNR by \citetads{1983IAUS..101..541M}. 
In X-rays, DEM L~299 has been observed and catalogued by the {\sl Einstein Observatory} \citepads{1981ApJ...248..925L,1991ApJ...374..475W} 
and \rosat\ \citepads{1999A&AS..139..277H}. 
The optical size of \deml\ was determined by \citetads{2010AJ....140..584D} 
from optical data of the Magellanic Clouds Emission Line Survey (MCELS) to be $5\farcm8
\times 4\farcm0$, which corresponds to $\sim$\,$84~\rm pc \times
58~\rm pc$, assuming a distance to the LMC of 50~kpc 
\citepads{2013IAUS..289..169P}, 
and has a central cavity of $1\farcm8 \times
2\farcm3$ ($\sim$\,$26~\rm pc \times 33~\rm pc$). With a diameter of 4\farcm3 ($\sim$\,$62~\rm pc$)
the X-ray dimensions are about the same as in the optical, while the radio size of
3\arcmin\ ($\sim$\,44~pc) is much smaller \citepads{1986A&A...166..257B}. 
At the north-western rim of \deml, there is a smaller \ion{H}{II} region with a
diameter of $\sim$\,1\arcmin\ that is not associated with the SNR. This contains
the young stellar object (YSO) and an OB-star \citepads[see][]{2010AJ....140..584D}. In the south-west of
\deml\ lies the bright \ion{H}{II} region \object{LHA~120-N~164}. 

Since we found evidence that the \hii\ region \deml\ also harbours a superbubble
in addition to the supernova remnant SNR B0543-68.9, we use the nomenclature
  '\deml' only for the entire \hii\ region and not for the remnant, which has previously also been labelled
  '\deml'.

The aim of this paper is to study this intriguing region using X-ray, optical, and
radio observations to investigate the varied morphology of \deml\ in the different
energy bands. 
In Sect.~\ref{Sect_Data} we describe the X-ray, optical, and radio
data we used for our studies and the analysis of these data. The morphological studies can be found in
Sect.~\ref{Sect_MorphologyStudies}, which contains comparative studies of X-ray, radio, and optical
data. The X-ray spectral analysis of both objects is presented in
Sect.~\ref{Sect_SpectralAnalysis}, with a discussion of our results given in
Sect.~\ref{Sect_Discussion}. In Sect.~\ref{Sect_Summary}, we summarise
our work and draw conclusions.


\begin{table*}
  \caption{Overview of the \xmm\ X-ray observations of \deml\ and the
    South Ecliptic Pole. The observations are background-flare filtered.}
  \label{TabObsID}      
  \centering                          
  \begin{tabular}{c c c c c c c c}       
    \hline\hline                 
   Observed Region & Obs.-ID    & Obs.-Date & PI         & \multicolumn{3}{c}{Eff. Obs. Time for EPIC}   & Filter\\    
               & & &            & MOS1 & MOS2 & pn            & \\
               & & &            & [ks] & [ks] & [ks]          &  \\
 \hline                                                          
    \deml & 0094410101 & 19.10.2001 & Y.-H.~Chu  & 11.2 & 11.3 & 7.9 &         Medium \\                      
    South Ecliptic Pole & 0162160101 & 24.11.2003 & B.~Altieri & 11.9 & 12.3 & 8.6  &         Thin1  \\
    South Ecliptic Pole & 0162160301 & 05.12.2003 & B.~Altieri & \phantom{0}8.5 & \phantom{0}8.5 & 7.1    &         Thin1  \\
    South Ecliptic Pole & 0162160501 & 14.12.2003 & B.~Altieri & \phantom{0}9.4 & \phantom{0}9.2 & 7.0    &         Thin1  \\
    \hline                                 
  \end{tabular}
\end{table*}

\section{Data}
\label{Sect_Data}


\begin{figure*}
    \includegraphics[width=0.5\textwidth]{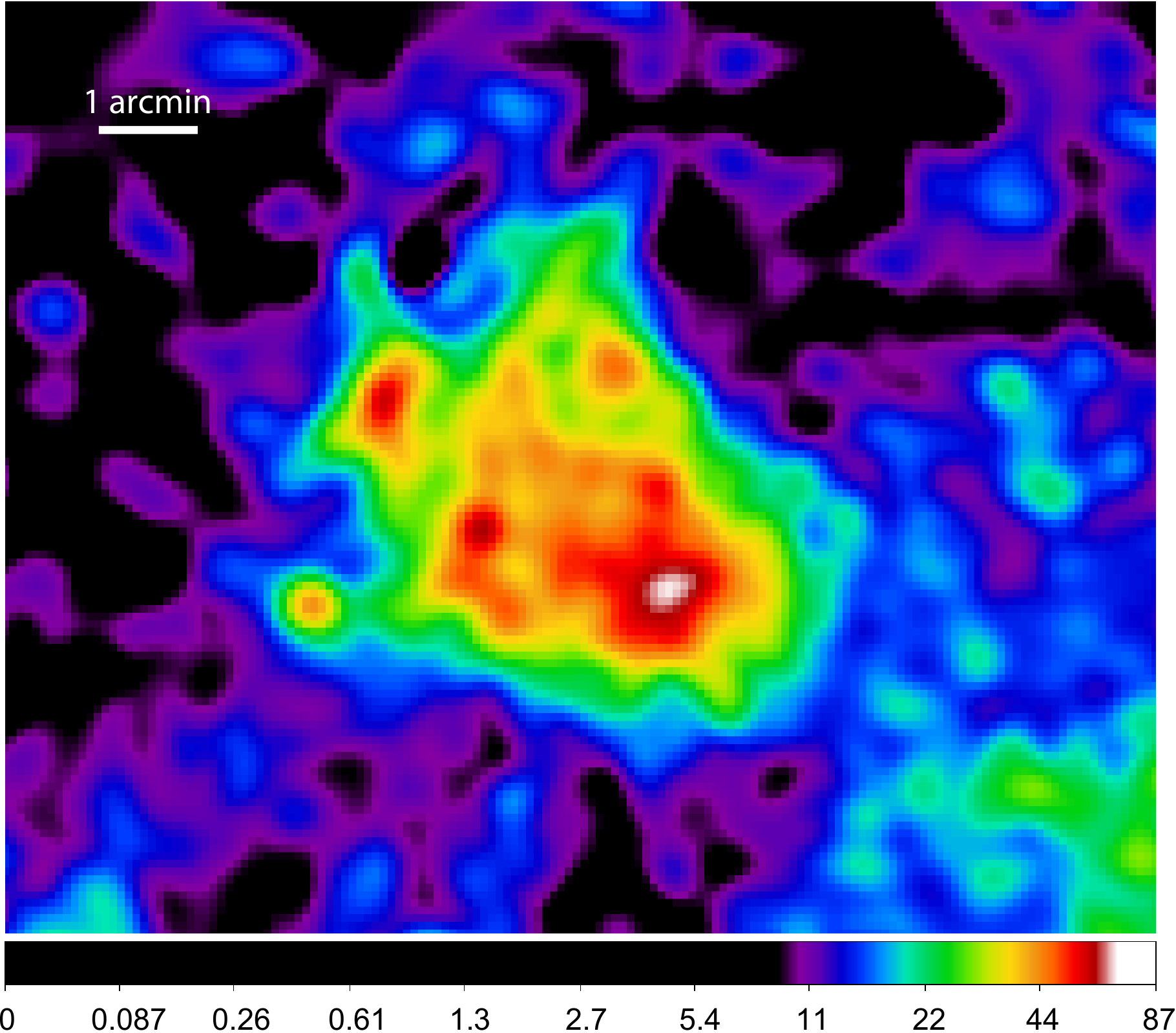}
  \includegraphics[width=0.5\textwidth]{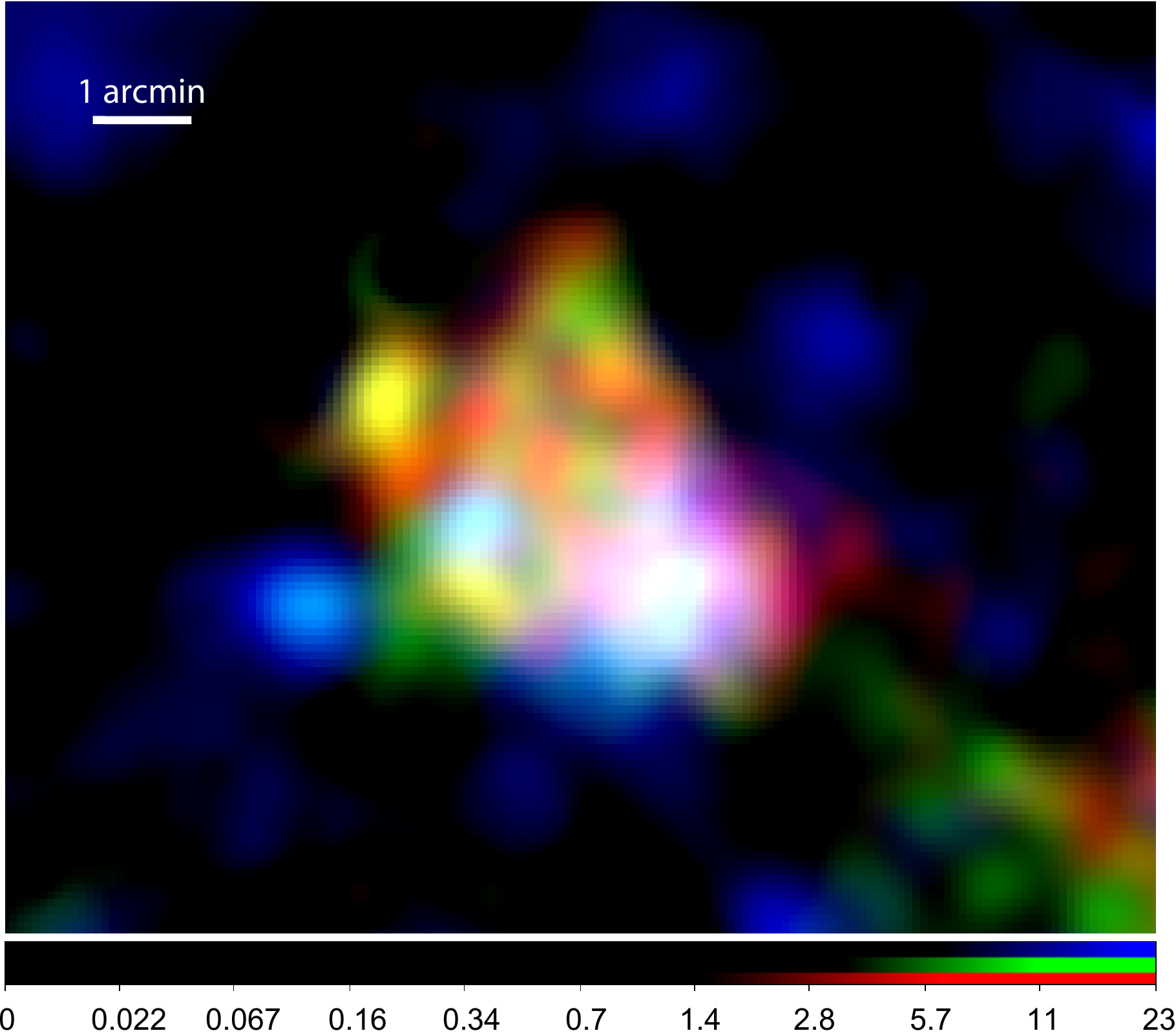}
  \caption{Left: intensity-scaled broadband \xmm\ X-ray image of \deml. The image shows an area of
    $12\farcm4 \times 10\arcmin\ $ in the energy range of 0.3--8.0~keV and has been smoothed by a
    Gaussian with a kernel radius of four pixels, after being adaptively
    smoothed.
    Right: three-colour image created out of \xmm\ data. Red: 0.3--0.8~keV,
     green: 0.8--1.5~keV, blue: 1.5--4.5~keV. To create the image,
     the same area, the same kernel radius, and the same scale units were used
     as for the broadband image. The scales are in counts/s/deg$^2$.}
         \label{FigAdaptXray}
   \end{figure*}

\subsection{X-ray data}
\label{Sect_Data_Xray}
For the X-ray studies, we analysed the archived data of the European Photon Imaging Camera
(EPIC) of the ESA satellite \xmm\ \citepads{2001A&A...365L...1J} 
with Obs.-ID 0094410101 (PI: Y.-H.~Chu) from the year 2001. The observation was performed in full-frame mode and has an effective observation time of $\sim$\,11~ks for the
EPIC MOS1 and EPIC MOS2 detectors \citepads{2001A&A...365L..27T}, 
and $\sim$\,8~ks for the EPIC pn detector
\citepads{2001A&A...365L..18S}. 
Table~\ref{TabObsID} summarises the observations.

We processed the data with the \xmm\ Extended Source Analysis Software
(XMM-ESAS) package version 4.3\footnote{ESAS:
\href{http://xmm.esac.esa.int/sas/current/doc/esas/index.html}{http://xmm.esac.esa.int/sas/current/doc/esas/index.html};
  \href{ftp://xmm.esac.esa.int/pub/xmm-esas/xmm-esas_4.3.pdf}{ftp://xmm.esac.esa.int/pub/xmm-esas/xmm-esas$\_$4.3.pdf}\label{FootESAS}} 
\citepads{2011AAS...21734417S,ESAScookbook}, 
which is part of the \xmm\ Scientific Analysis Software version 11.0.0\footnote{SAS: \href{http://xmm.esa.int/sas/}{http://xmm.esa.int/sas/}} (SAS). We filtered out soft proton flares by light-curve screening, removed point sources,
subtracted the quiescent-particle background using filter-wheel-closed data
and unexposed corners of the detectors, and checked for EPIC MOS CCDs operating in an
anomalous state. This is a state that shows an enhanced background in the energy
range $\la 1$~keV
\citepads{2008A&A...478..575K}
. MOS1 CCD\#5 was identified to be in this state and was therefore excluded from
our analysis. More details about these ESAS procedures can be found in
\citetads{2008A&A...478..615S}.

To estimate the cosmic X-ray background (CXB) for the spectral analysis, we
used three archived \xmm\ EPIC
observations of the South Ecliptic Pole
(SEP) from 2003 with Obs.-IDs 0162160101, 0162160301, and 0162160501 (PI:
B.~Altieri). 
We assumed that the CXB of the
SEP observations is representative of the \deml\ observation. All three SEP observations have been taken in full-frame mode and result in a total
effective observation time of $\sim$\,30~ks (see Table
\ref{TabObsID}). The data of these observations were analysed in the
same way as described above. For Obs.-ID 0162160301, MOS1 CCD\#4 was found to be operating in
an anomalous state and thus excluded from further
analysis. The resulting data sets of these three
observations were combined to
obtain one dataset with an effective exposure time of 30~ks for the MOS1 and
MOS2 detectors, and 23~ks for the pn detector. 

Images of \deml\ created out of the \xmm\ data are shown in Fig.~\ref{FigAdaptXray} (see
Sect.~\ref{Sect_MorphologyStudies} for details).


\subsection{Optical data}

For our optical studies, we used H$\alpha$, [\ion{S}{II}], and [\ion{O}{III}] data 
of the MCELS\footnote{MCELS:
  \href{http://www.ctio.noao.edu/mcels/}{http://www.ctio.noao.edu/mcels/}}
\citepads[e.g.][]{2000ASPC..221...83S}, 
which mapped the Large and the Small Magellanic Clouds in \halpha\ (6563~\AA), [\ion{S}{II}] (6724~\AA), [\ion{O}{III}]
(5007~\AA), and at two continuum bands, which were used to subtract the stellar continuum emission. The survey was carried out by the University of Michigan (UM) and the Cerro Tololo
Inter-American Observatory (CTIO) and was performed with the UM/CTIO
Curtis Schmidt telescope situated in Chile. This telescope is a
0.6~m Schmidt telescope with a resolution of 2\farcs3/pixel. Its camera is a
SITE $2048 \times 2048$ CCD and produces images with a field of view of $1.35^{\circ} \times
1.35^{\circ}$ per image. The data we used for our studies were sky-subtracted, flux-calibrated, and continuum-subtracted. For the smoothed images created out of these data, the exposure time was normalised to one second.


\subsection{Radio-continuum data}

\begin{figure*}
 \includegraphics[trim=0.0cm 1.6cm 0.0cm 0.0cm, clip=true,width=0.5\textwidth]{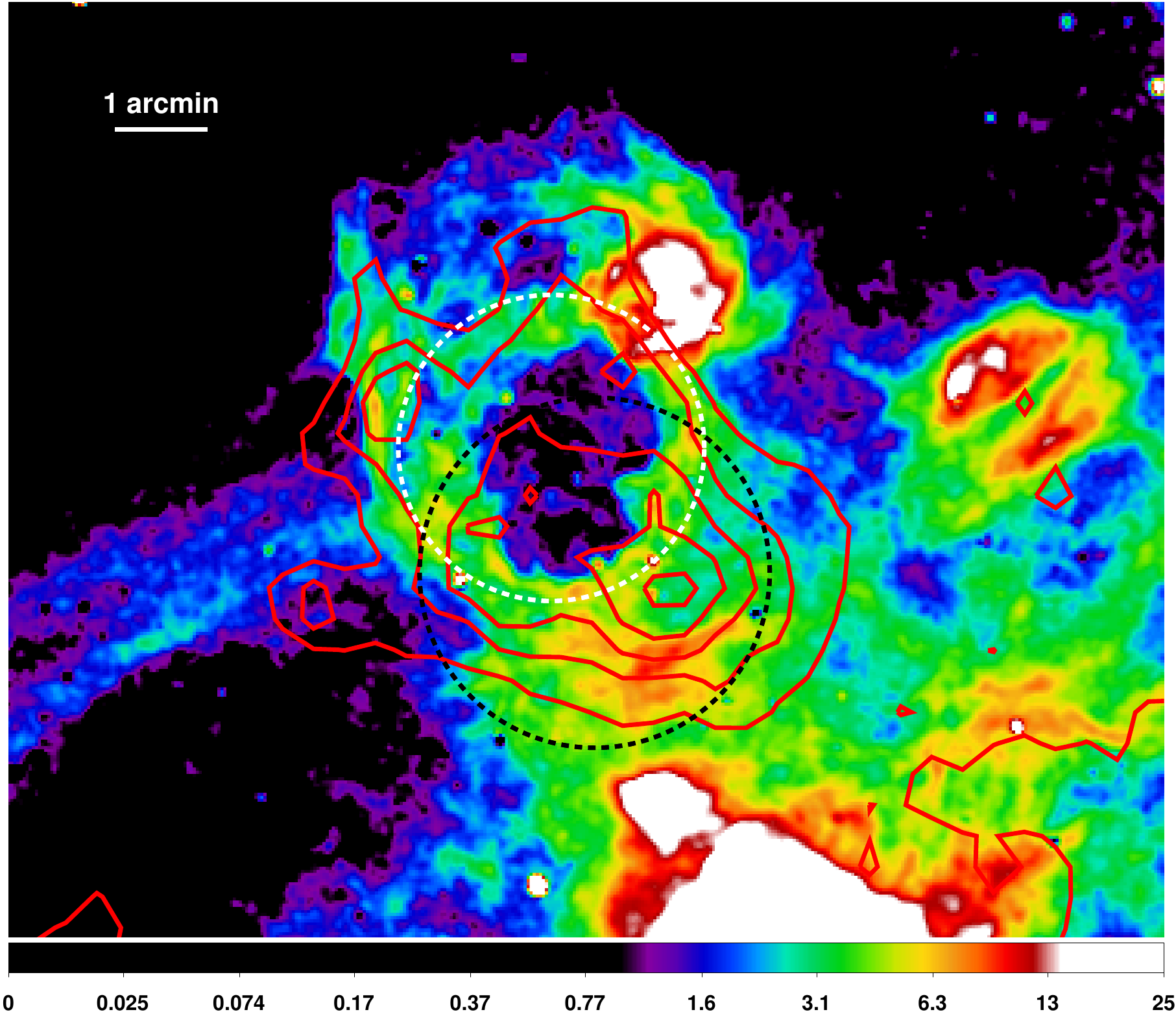}
 \includegraphics[trim=0.0cm 1.6cm 0.0cm 0.0cm, clip=true,width=0.5\textwidth]{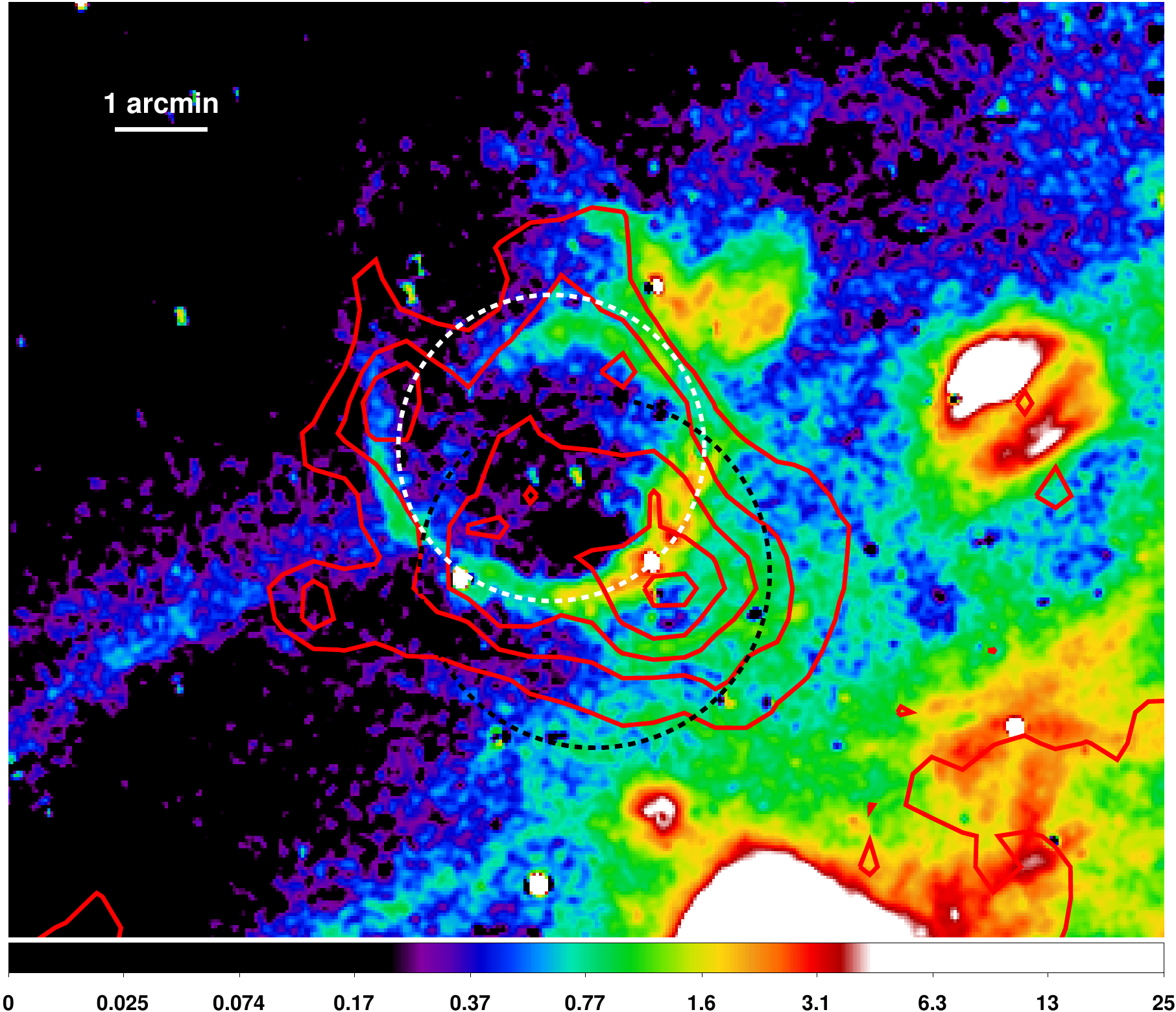}
 \caption{MCELS images in H$\alpha$ (left) and
       [\ion{O}{III}] (right). The images are continuum-subtracted,
       flux-calibrated, sky-subtracted, and smoothed. 
The same area of the sky is shown as in the X-ray images in Fig.~\ref{FigAdaptXray}. The contours are taken from the smoothed X-ray broadband image, with 5 contour
       levels at 18, 28.5, 39, 49.5, and 60~counts/s/deg$^2$. The white dashed circle
       shows the position of the \halpha\ cavity, and the black dashed circle the
       position of the shell in the \siihalpha\ image (see Fig.~\ref{FigOpticalPointsSDivH}). 
}
\label{FigOpticalPointsHO}
\end{figure*}

\begin{figure}
\includegraphics[width=0.5\textwidth]{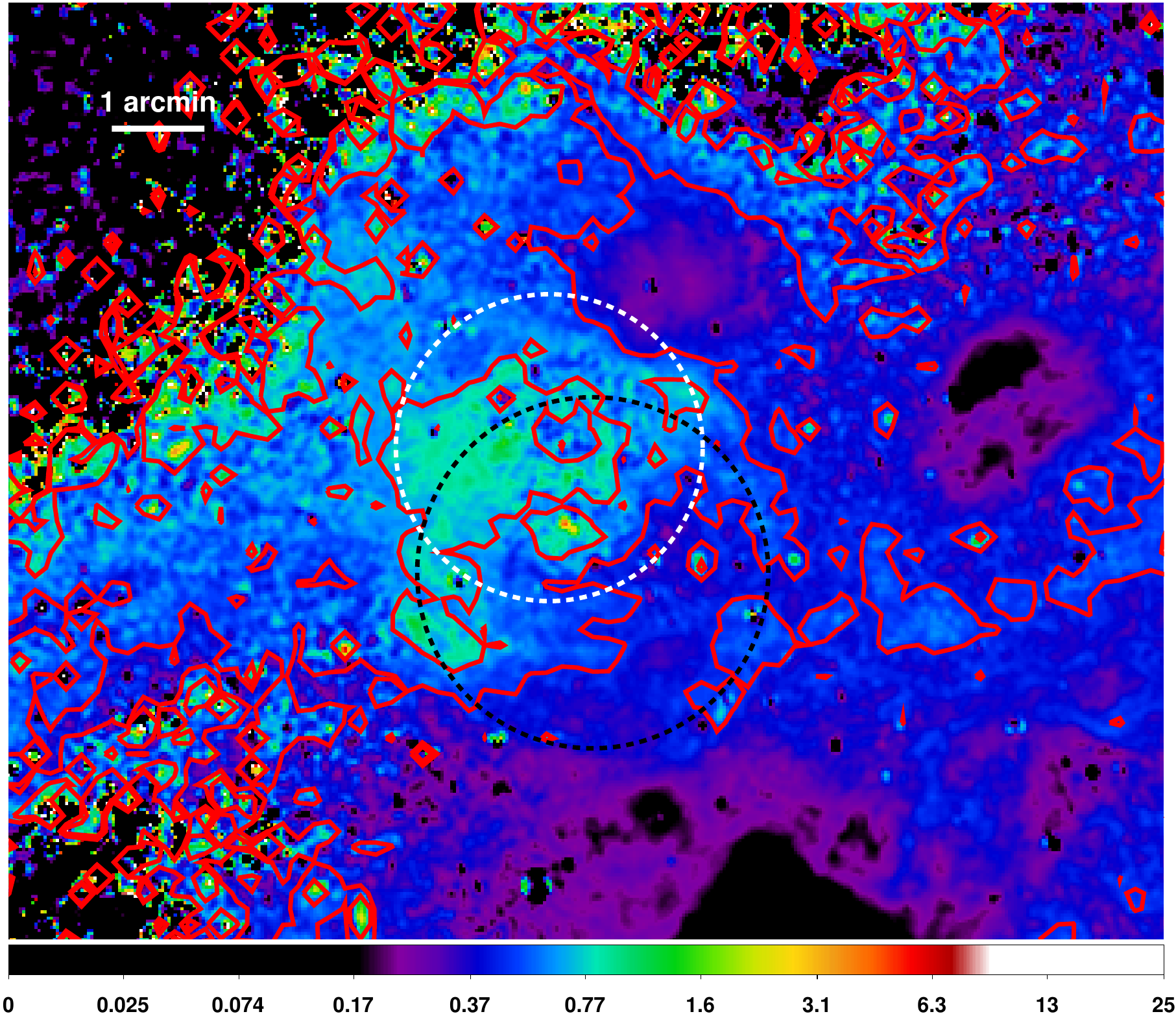}
  \caption{[\ion{S}{II}]/H$\alpha$ flux-ratio image created out of
    the flux-calibrated,
    continuum-subtracted 
    MCELS H$\alpha$ and [\ion{S}{II}] data (see
    Fig.~\ref{FigOpticalPointsHO}). A circular structure is
    visible, indicated by the black dashed circle. The north-eastern part of this
    structure has a \siihalpha\ flux ratio greater than
  0.67 and thus seems to be shock-ionised. Therefore, this structure is indicative of an SNR.
The position of the
  \halpha\ cavity is marked
  in white. The flux ratio contours are shown with contour levels at
  0.46 and 0.67. The area of the sky is the same as in
  Fig.~\ref{FigOpticalPointsHO}.
 }
  \label{FigOpticalPointsSDivH}
\end{figure}
\begin{figure}
    \includegraphics[width=0.505\textwidth]{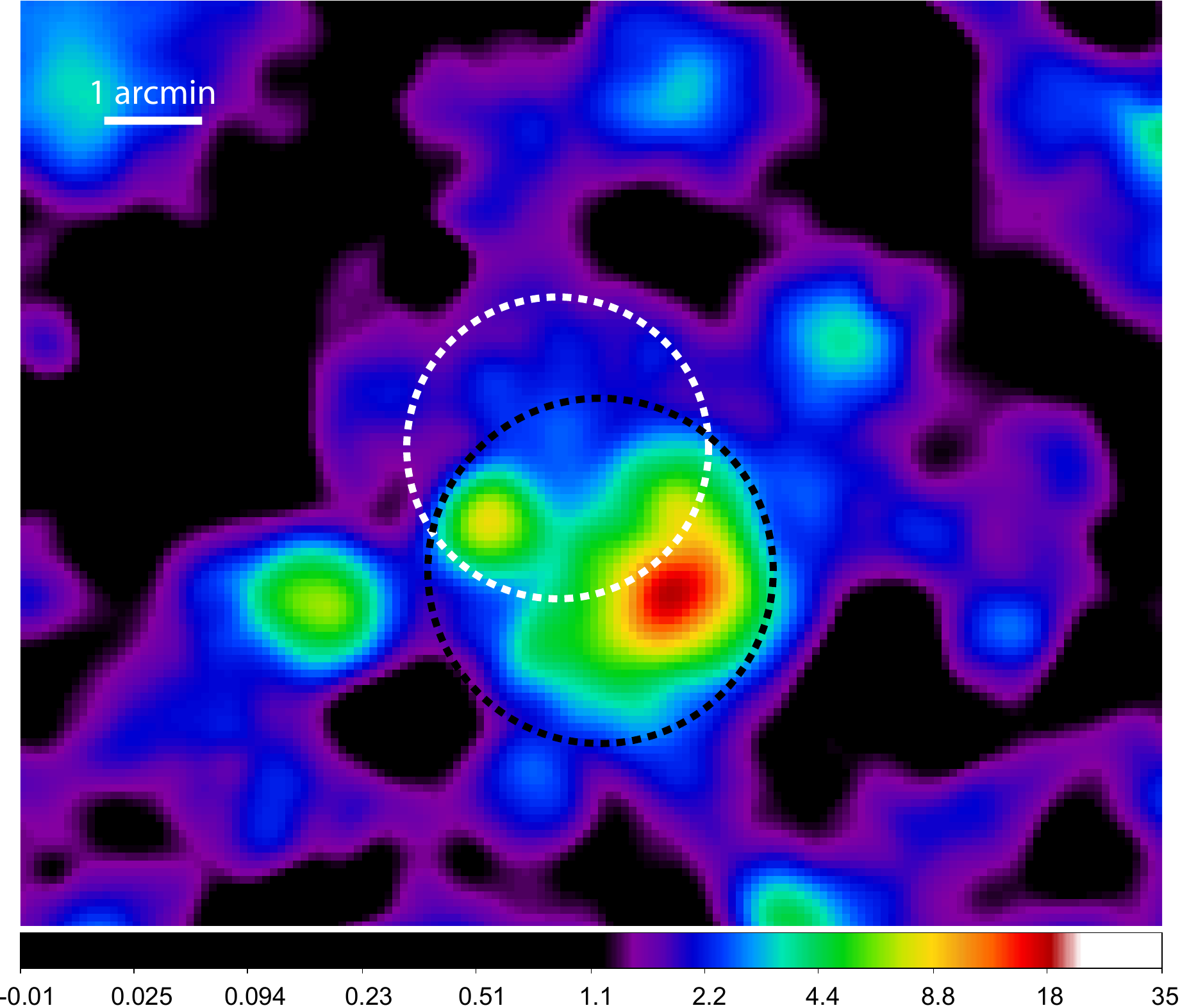}
  \caption{Intensity-scaled X-ray image of \deml\ in an energy range
    of 1.5--4.5~keV. The black dashed circle indicates the
    position and size of the SNR as indicated through
        the \siihalpha\ flux ratio image (Fig.~\ref{FigOpticalPointsSDivH}), 
    and the white dashed circle the
    superbubble as indicated through the optical images
    (Fig.~\ref{FigOpticalPointsHO}). Most of the hard X-ray emission is confined inside the
    SNR region. Area of the sky, scale units, and smoothing parameters are the same as in Fig.~\ref{FigAdaptXray}.}
         \label{FigHardXray}
   \end{figure}

We used radio observations at four frequencies (see Table~\ref{tbl-1}) to study
and measure flux densities of \deml. For the 36~cm (Molonglo Synthesis
Telescope, MOST) flux density measurement given in Table~\ref{tbl-1} we used
unpublished images as described by \citetads{1984AuJPh..37..321M}, and for
the 20~cm we used image from \citetads{2007MNRAS.382..543H}. The Australia
Telescope Compact Array (ATCA) project C634 (at 6/3~cm) observations were
combined with mosaic observations from project C918
\citepads{2005AJ....129..790D}. Data for project C634 were taken by the ATCA on
1997 October~6/7 and 17/18, using the array configuration EW375 and 750C. For
the final image (Stokes parameter \textit{I}) we excluded baselines created
with the $6^\mathrm{th}$ ATCA antenna, leaving the remaining five antennas to
be arranged in a compact configuration. C634 observations were carried
out in snap-shot mode, totalling $\sim$\,4.5~hours of integration over two
12~hour periods. Source PKS~B1934--638 was used as the primary calibrator and
source PKS~B0530--727 as the secondary calibrator. The
\textsc{miriad} \citepads{2006Miriad} 
and \textsc{karma} \citepads{1996ASPC..101...80G} 
software packages were used for reduction and analysis. The 6~cm and 3~cm images
were constructed using \textsc{miriad} multi-frequency
synthesis \citepads{1994A&AS..108..585S}. Deconvolution was achieved with the
{\sc clean} and {\sc restor} tasks with primary-beam correction applied using
the {\sc linmos} task. Similar procedures were used for the \textit{U} and
\textit{Q} Stokes parameters. Using the flux density measurements shown in Table~\ref{tbl-1}, we estimated a radio spectral index for \deml\ of
$\alpha=-0.34 \pm 0.03$. More information on the observing procedure and
other sources observed in this session/project can be found in
\citetads{2007MNRAS.378.1237B}, 
\citetads{2008SerAJ.177...61C,2008SerAJ.176...59C,2010A&A...518A..35C}, 
\citetads{2009SerAJ.179...55C}, 
\citetads{2012A&A...540A..25D}, 
\citetads{2012A&A...539A..15G}, 
\citetads{2012A&A...543A.154H}, 
\citetads{2012A&A...546A.109M}, 
\citetads{2013A&A...549A..99K}, and 
\citetads{2010SerAJ.181...43B,2012SerAJ.184...69B,2012MNRAS.420.2588B,2012RMxAA..48...41B,2012SerAJ.185...25B,2013MNRAS.432.2177B}.



\begin{figure*}
  \includegraphics[width=0.5\textwidth]{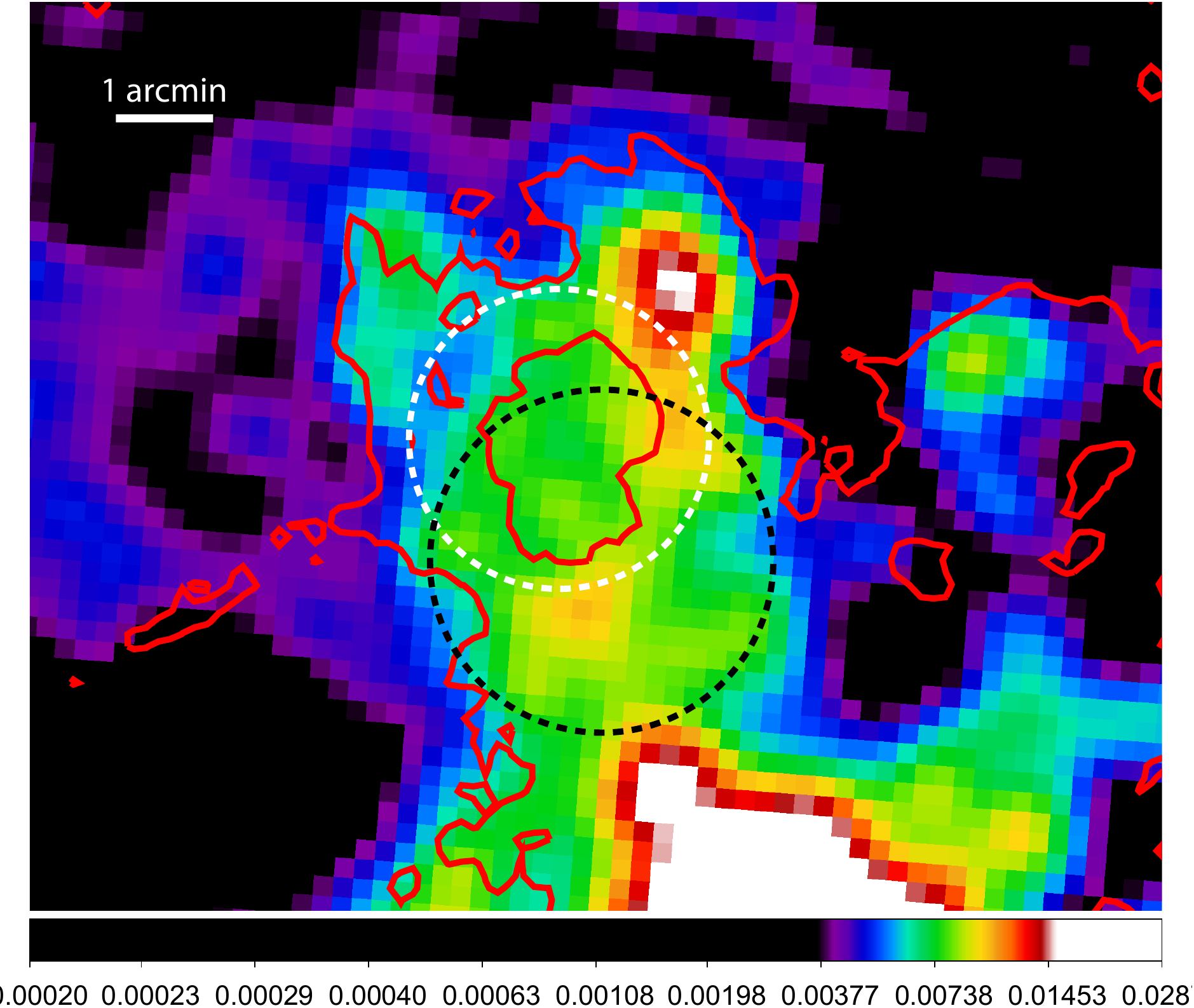}
 \includegraphics[width=0.5\textwidth]{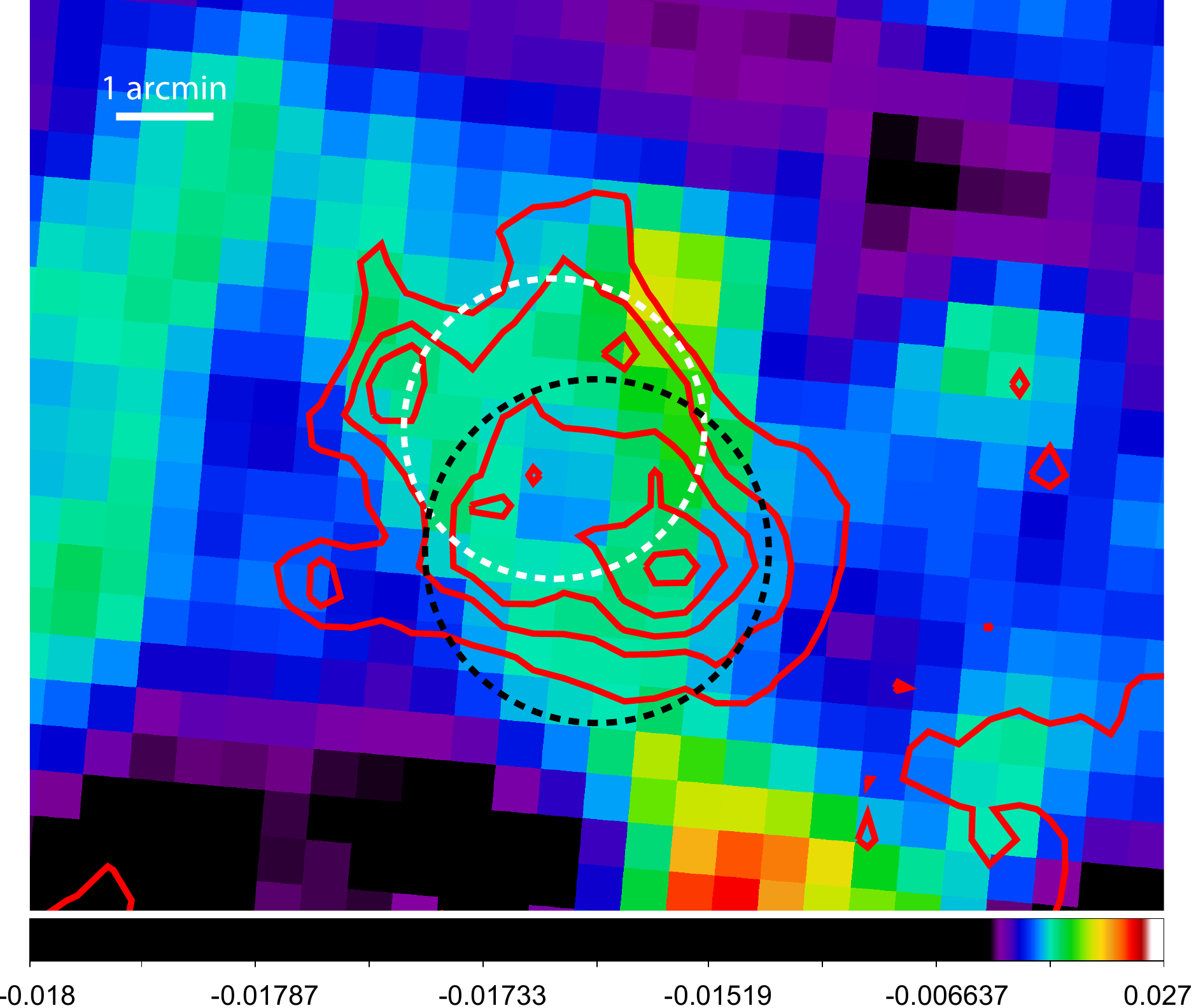}
 \caption{Left: 20 cm radio-continuum image of \deml\ with overlaid \halpha\
    contours, showing the same area of $12\farcm4 \times 10\arcmin\ $
    as the X-ray images in Fig.~\ref{FigAdaptXray}. The
    image was smoothed with a Gaussian using a kernel radius of one pixel. The
    contours are taken from the \halpha\ image in Fig.~\ref{FigOpticalPointsHO} with one contour
    level at 2. The black dashed circle indicates the position of the SNR, while the
    white dashed circle shows the position of the superbubble. The scales are
    in Jy/beam. Right: same as left, but for the 36 cm radio-continuum data and with the same X-ray contours
    as in Fig.~\ref{FigOpticalPointsHO}.}
         \label{FigRadioFilipovich}
\end{figure*}

\section{Morphology studies}
\label{Sect_MorphologyStudies}


\subsection{X-rays}

To create smoothed, exposure-corrected 
images in different energy bands, we combined the data of the MOS1, MOS2, and pn detectors
using ESAS, rebinned the data with a factor of two, and adaptively smoothed it with
a smoothing counts value of 50 using the ESAS task \textit{adapt\_900}. Afterwards, the images were smoothed
again with a Gaussian using a smoothing kernel radius of four pixels for a
better presentation.

A zoom-in on the \deml\ region is shown
on the left side of Fig.~\ref{FigAdaptXray} in the energy
range of 0.3--8.0~keV (broadband). We furthermore created images in energy bands of
0.3--0.8~keV (soft band), 0.8--1.5~keV (medium band), and 1.5--4.5~keV
(hard band). These images were combined to produce the three-colour image on the right side of
Fig. \ref{FigAdaptXray}. Both images show the same area of the sky with a size of
$12\farcm4 \times 10\arcmin$, which corresponds to $\sim$\,180~$\rm pc
\times 145~\rm pc$. At the northern rim of \deml, a point-source was excluded that 
correlates with the position of the stellar X-ray source
\object{[SHP2000] LMC\,347} \citepads{2000A&AS..143..391}. 
The highest X-ray emission is reached in the south-west (see Fig.~\ref{FigAdaptXray}, left). This part is bright in all three energy bands, as can be seen in the
three-colour image on the right side of Fig.~\ref{FigAdaptXray}, where it
appears white in colour. Since the emission around the brightest region
declines in all three energy bands and not only in the soft energy band, the
difference in brightness does not seem to be an effect of absorption.


\subsection{Optical}

\begin{figure}
  \includegraphics[trim=0.0cm 0.0cm 0.0cm 1.5cm, clip=true,angle=270,width=0.5\textwidth]{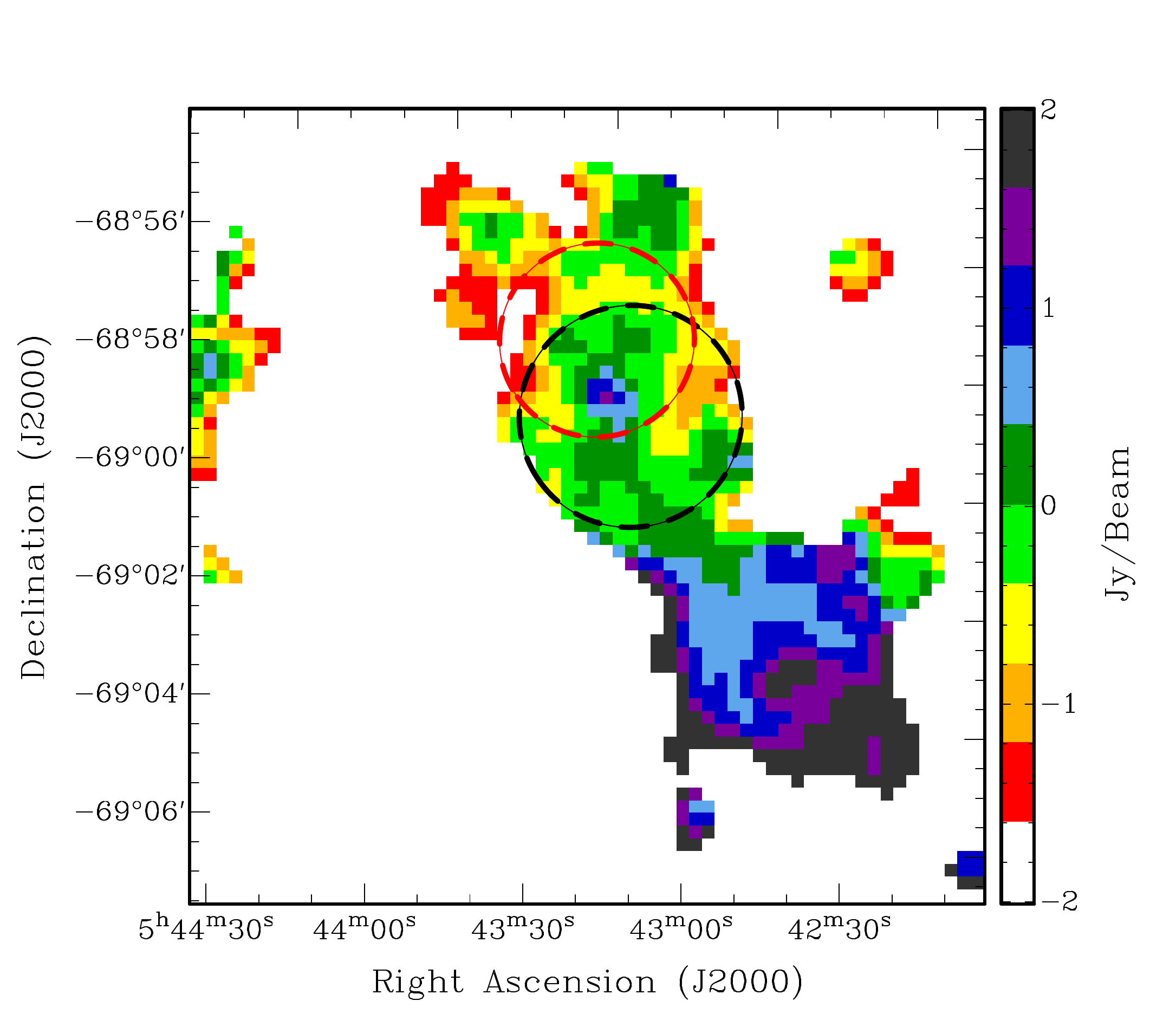}
 \caption{Spectral index map of the \deml\ region created out of the 20~cm and
 36~cm images. The position of the SNR and of the superbubble are marked in
 black and red (white in other figures). \deml\ can be easily distinguished from the
 \ion{H}{II} region LHA 120-N 164, which lies to the south-west of it,
 because of their different spectral indices. }
         \label{FigRadioSpectralIndex}
\end{figure}
\begin{figure}
  \includegraphics[width=0.49\textwidth]{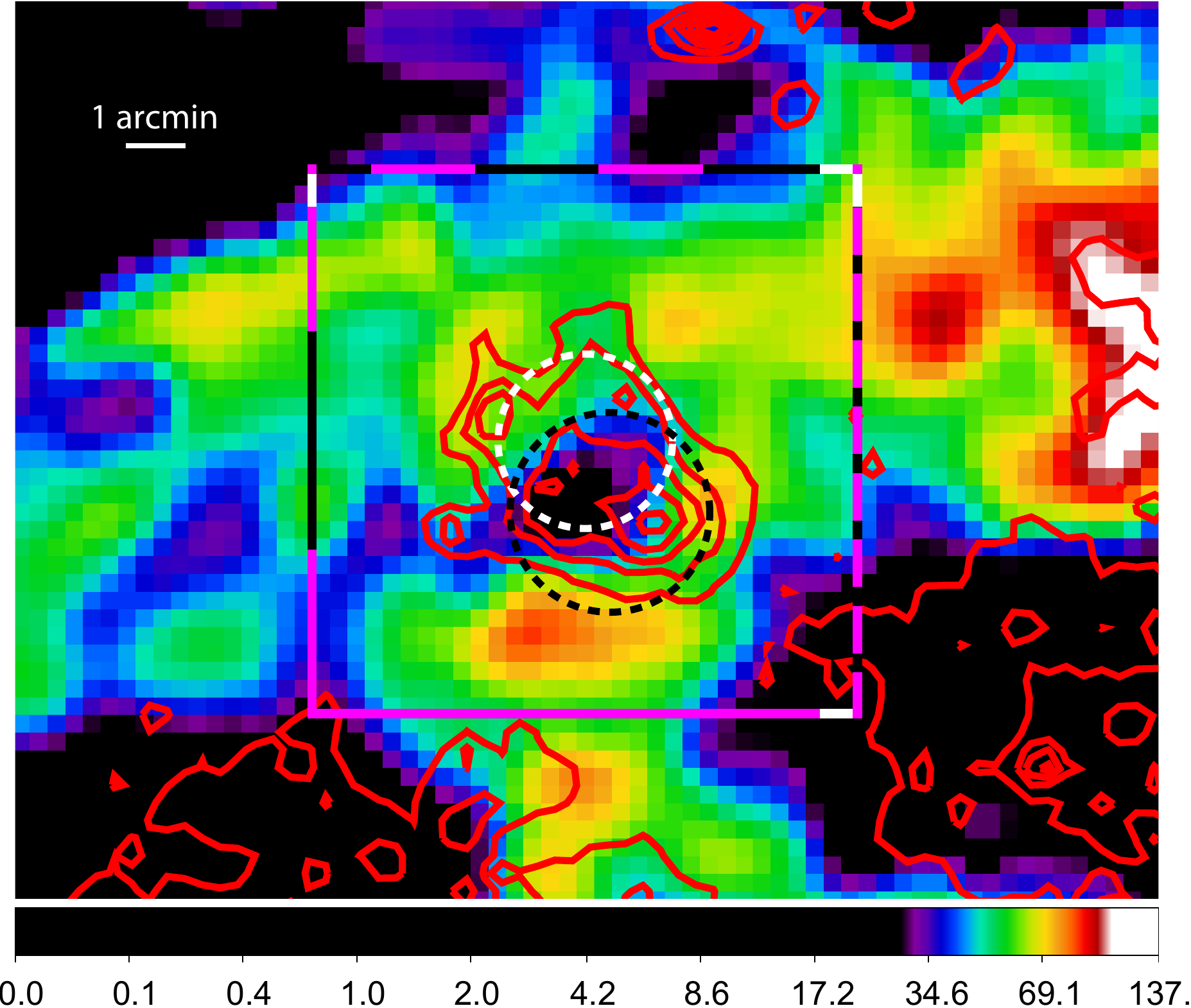}
  \caption{Combined ATCA and Parkes \ion{H}{I} 21~cm radio data showing
    a velocity slice at $\sim$\,264~km/s in the
    region of $22\arcmin\ \times 17\arcmin\ $ around \deml\ with
    overlaid X-ray contours of Fig.~\ref{FigAdaptXray}. Scales represent the
    brightness temperature and are in Kelvin. The data have been
    smoothed by a Gaussian using a kernel radius of two pixels. The black dashed circle indicates the approximate size and position of the SNR based on 
    the [\ion{S}{II}]/H$\alpha$ ratio, while the white dashed circle marks the
    position of the H$\alpha$ cavity. The 
    box indicates the
    area that we used to create the images in Fig.~\ref{FigRadioXY} at
    different heliocentric velocities and the position-velocity images shown in
    Fig.~\ref{FigRadioPV}. The different line styles of the box mark its
    individual sides and their orientation (white mark), for an easier comparison with the
    plots shown in Fig.~\ref{FigRadioPV}. 
   }
   \label{FigRadio21cm}
\end{figure}

We created sky-subtracted, flux-calibrated, continuum-subtracted,
and smoothed images from the MCELS data. These
images can be seen in Fig.~\ref{FigOpticalPointsHO} and show the same area
of the sky as in Fig.~\ref{FigAdaptXray}
in H$\alpha$ (left side) and [\ion{O}{III}] (right side), with overlaid X-ray
contours.

These images reveal a completely different appearance of \deml\ in the optical
than in X-rays. The emission line images are dominated by a circular shell-like
structure, which is about 4\arcmin\ ($\sim$\,60~pc) in diameter and possess a
visible cavity with a diameter of $\sim$\,2\farcm5 ($\sim$\,35~pc). Compared with the X-ray
image, the optical cavity lies in the eastern, fainter
part of the X-ray emission, and does not include the brightest area of X-ray
emission in the south-west. The border of the supergiant shell LMC-SGS~2 can be seen as a filament east of \deml.
While the optical shell is quite clearly defined in the west and the east (cf. [\ion{O}{iii}], Fig.~\ref{FigOpticalPointsHO},
right side), the north-north-eastern border shows less emission and is much more diffuse than the rest of the shell-like
structure. The [\ion{S}{II}] emission (not shown here) bears a strong resemblance to
the H$\alpha$ emission (Fig.~\ref{FigOpticalPointsHO},
left). The [\ion{O}{iii}] emission (Fig.~\ref{FigOpticalPointsHO}, right) shows some distinct
differences than the \halpha\ and [\ion{S}{II}]
emission. The shell of the cavity is much
narrower with a very circular shape. There is a break in this shell in the
north-east where the \halpha\ and \sii\ emission show a
diffuse structure (see Fig.~\ref{FigOpticalPointsHO}), indicating a
possible blowout. This suggestion 
is supported by a slight enhancement of the X-ray
emission in this region, which is discussed in
Sect.~\ref{Sect_SpectralAnalysis}. The main
difference between these optical images is the lack of [\ion{O}{iii}] emission in the
south-west below the cavity, which is located just south of the
brightest region in X-rays.

\begin{figure*}
  \includegraphics[width=\textwidth]{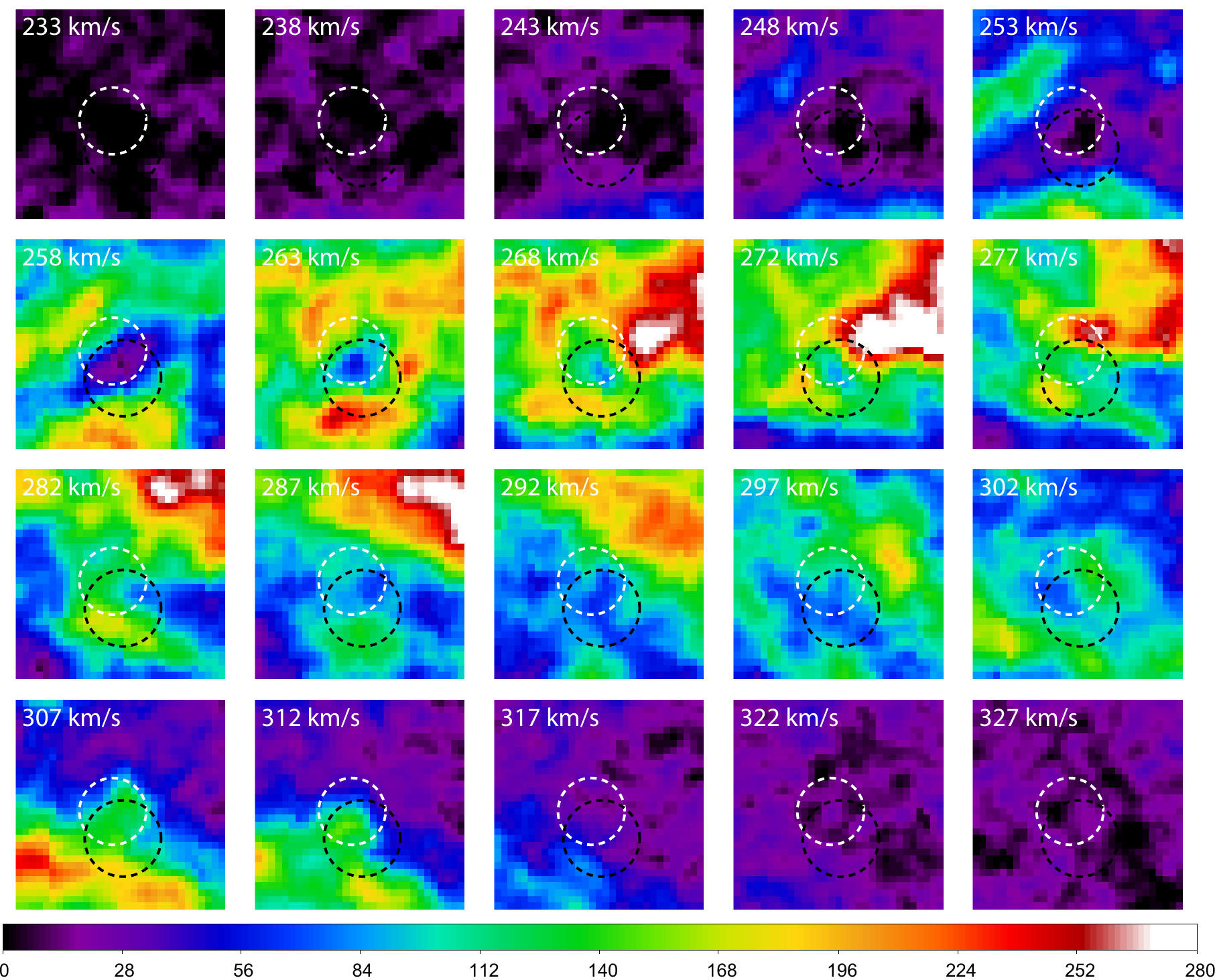}
  \caption{\ion{H}{I} data cube channel maps showing
    the \ion{H}{I} distribution for a
    certain heliocentric velocity interval. Each plot is created out of three velocity
    slices, corresponding to a velocity interval of $\sim$\,5~km/s from top left to
    bottom right. The upper left
    plot starts with a mean heliocentric velocity of $\sim$\,233~km/s, the last plot in the lower right
    with $\sim$\,327~km/s. The black dashed circle marks the position of the SNR, while the
    white dashed circle corresponds to the H$\alpha$ cavity. Each image shows an
    area of $10\farcm3 \times 10\farcm3$, which is also indicated by the
    box in Fig.~\ref{FigRadio21cm}. Scales represent the
      brightness temperature and are in Kelvin.}
  \label{FigRadioXY}
\end{figure*}

\begin{figure*}
  \includegraphics[width=0.5\textwidth]{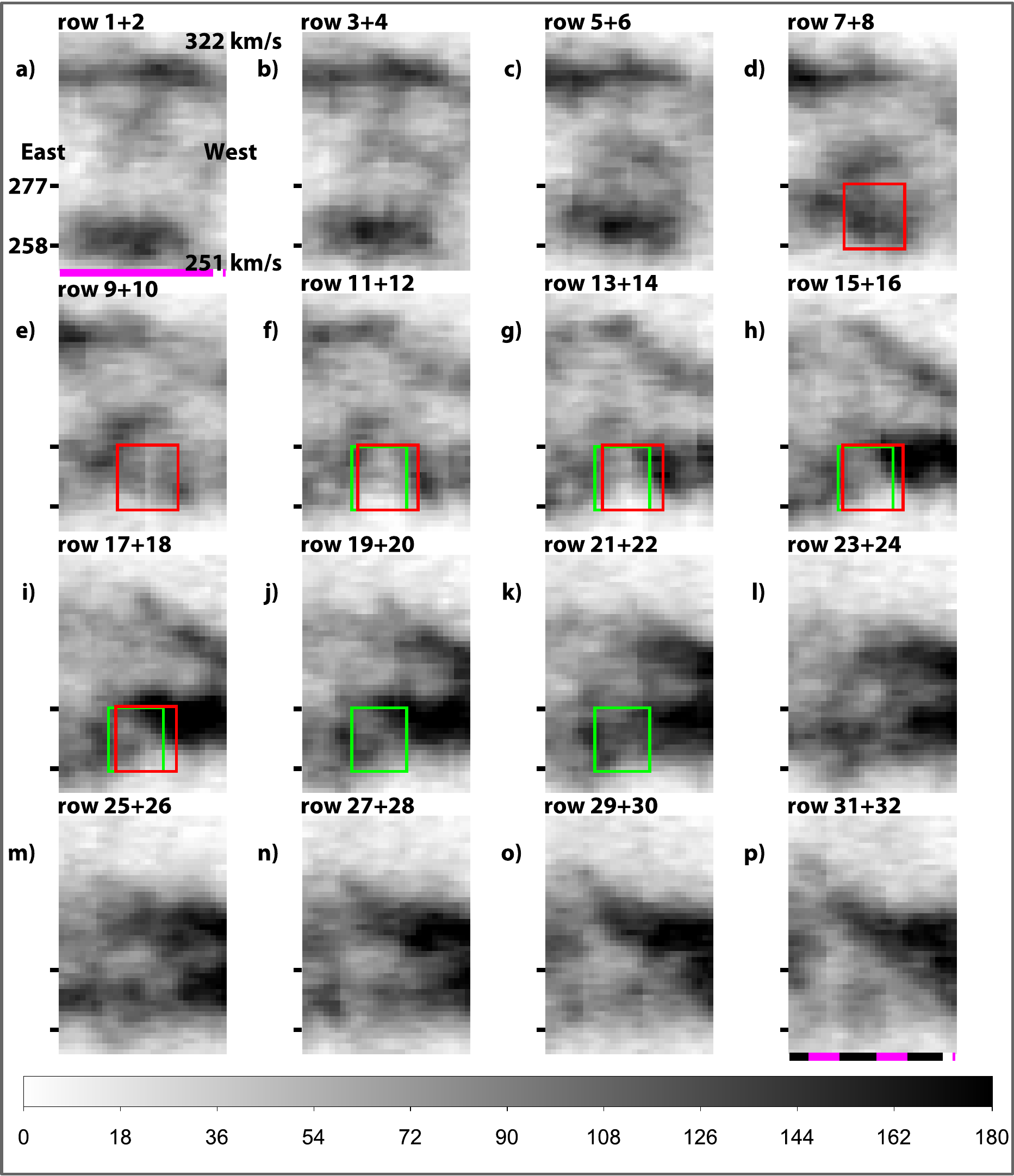}
   \includegraphics[width=0.5\textwidth]{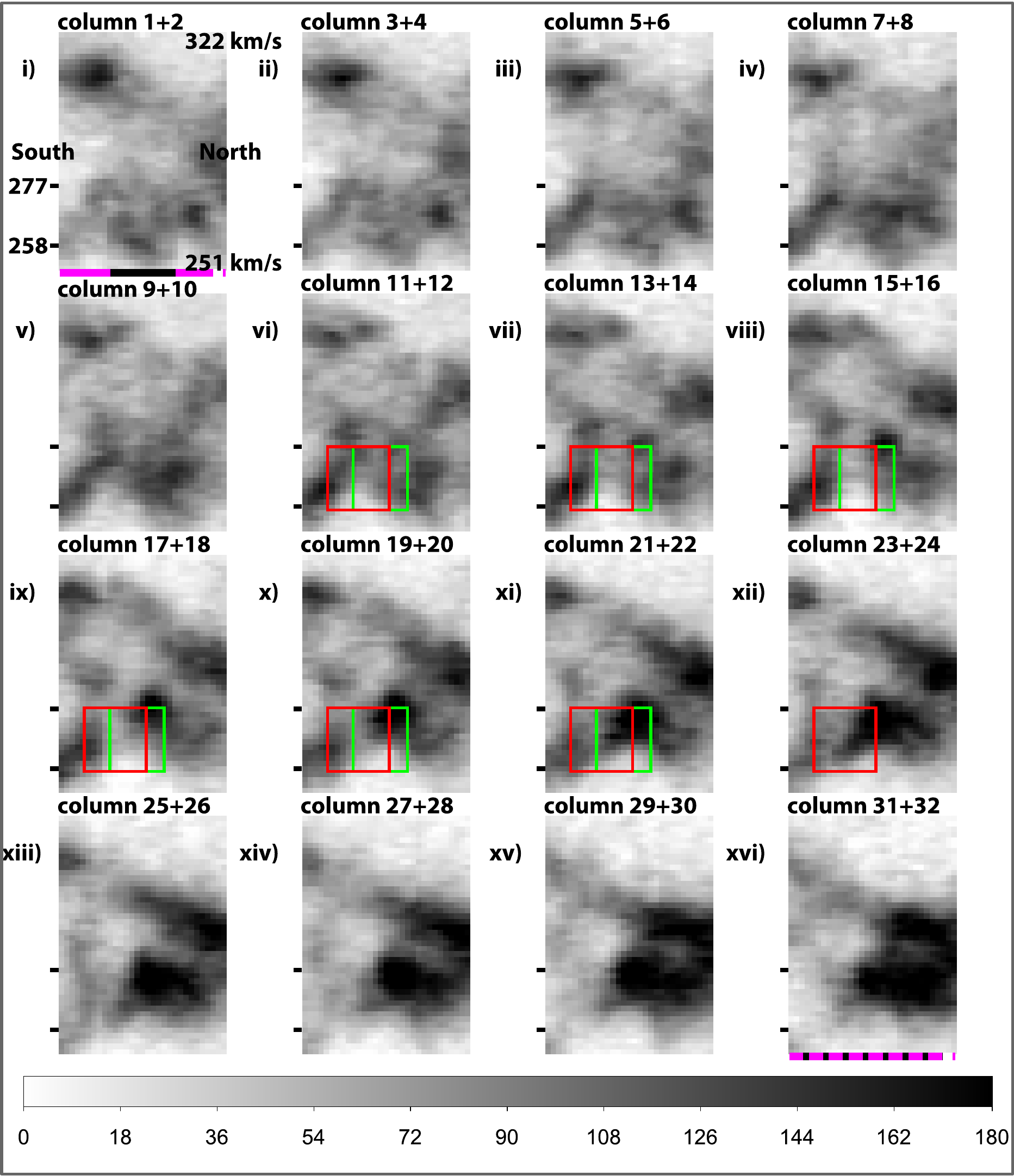}
 \caption{Average \hi\ distribution image of the \deml\ region at a certain position (left:
     east-west direction, right: south-north direction) for a velocity
     range of 251--322~km/s. Left: position-velocity plot showing
     the intensity of the \hi\ emission in dependence of the velocity (ordinate) of the gas within the box in Fig.~\ref{FigRadio21cm} against
   its position (abscissa)
   in the east-west direction of that box. 
The ordinate shows the velocity in the range of
251~km/s (at the bottom of each plot) to 322~km/s (at the top of each plot),
with a velocity resolution of 1.649~km/s. The abscissa
covers a range of 10\farcm3 in east-west
direction, corresponding to $\sim$\,150~pc at
the LMC distance.  
The different plots show from a) to p)
    the velocity information 
for two added pixel
    rows of the box in Fig.~\ref{FigRadio21cm}, as indicated above each plot. These pixel rows are taken from the box of Fig.~\ref{FigRadio21cm} from
    bottom to top. Since the abscissa of plot a) corresponds to the
      lowest two pixel rows of the box in Fig.~\ref{FigRadio21cm}, both are
      marked with a magenta line of the same line style (here: solid), also indicating their orientation
      through a white bar, for an easier comparison of these plots with
      Fig.~\ref{FigRadio21cm}. The same holds for plot p). 
      The red rectangle indicates the position of the SNR as
    indicated by the [\ion{S}{II}]/H$\alpha$ flux ratio, while the green rectangle shows the
    position of the H$\alpha$ central cavity. The velocity range of these
    rectangles corresponds to 258--277~km/s, as determined for the SNR and the
    superbubble in Fig.~\ref{FigRadioXY}. This velocity
      range is marked in the plots by two black tick marks. Scales are in Kelvin. Right: same as left, but for the south-north direction
    of the box 
    in Fig.~\ref{FigRadio21cm}, i.e. for
   the different pixel columns of the box (stepped from left to right)
   instead of pixel rows. 
  }
   \label{FigRadioPV}
\end{figure*}

To determine the exact position of the SNR, we examined the flux ratio of
[\ion{S}{II}] and H$\alpha$. \citetads[][and references therein]{1985ApJ...292...29F} 
stated that a [\ion{S}{II}]/H$\alpha$ flux ratio of $\gtrsim$\,0.67 is a strong indication for the
shock-ionised shell of an SNR that is located in an \ion{H}{II} region. We
created a [\ion{S}{II}]/H$\alpha$ ratio image as shown in
Fig.~\ref{FigOpticalPointsSDivH}. In this
image, a clear circular structure with an enhanced [\ion{S}{II}]/H$\alpha$
ratio is visible (black dashed circle). This circular structure has its centre at
RA 05:43:02.2, Dec -69:00:00.0 (J2000) and
a radius of
$\sim$\,2\arcmin\ ($\sim$\,30~pc). Emission located along the north-eastern
rim of this structure shows a flux ratio
of $\geq 0.67$, which is
an indicator for shock-ionised gas. Most likely, this structure
represents the border of the SNR. Surprisingly,
this shell is not at the same position as the H$\alpha$ cavity (cf. Fig.~\ref{FigOpticalPointsSDivH},
left side). Its centre
lies $\sim$\,1\farcm4 more to the south and $\sim$\,0.5\arcmin\
more to the west than the centre of the cavity. In contrast, no shell-like
structure is visible in the [\ion{S}{II}]/H$\alpha$ image at the position of
the cavity in \halpha\ or [\ion{O}{III}]. This led us to conclude that we see a superposition of two separate objects: an SNR and a
superbubble. The superbubble lies north of the SNR and corresponds to the position of a
shell clearly visible in \oiii, with some overlap. The
south-western part of the SNR, which has a lower flux ratio, corresponds to the
position of the X-ray bright region, which is confined inside the borders of
the optical SNR (see Fig.~\ref{FigHardXray}).


\subsection{Radio-continuum}

The 20~cm and 36~cm radio-continuum data of the \deml\ region are shown in Fig.
\ref{FigRadioFilipovich}. The two images show the same area of the sky as the X-ray images in
Fig.~\ref{FigAdaptXray}. The 20~cm and 36~cm images show extended emission coincident with
the X-ray position of \deml. The emission follows the H$\alpha$
emission rather than the X-ray emission (see
Fig.~\ref{FigRadioFilipovich}, left). A shell-like structure can be
seen in the 36~cm image, corresponding to the position of the two objects. Figure~\ref{FigRadioSpectralIndex} shows the spectral index map of
\deml\ using the 36~cm and 20~cm images. This spectral index map allows a clear differentiation of the SNR from
the \ion{H}{II} region LHA 120-N 164, which lies south-west of it (see
Fig.\ref{FigHalphaBig}). 
However, our spectral index estimate for the entire SNR ($\alpha=-0.34$) based
on all measured flux densities is slightly on the flatter side,
indicating the dominance of the thermal emission and the therefore somewhat older age of the
remnant. We also point out that the \ion{H}{II} region LHA 120-N 164
shows the canonical values of $\alpha \approx\,$+0.4.

  \begin{table}
  \centering   
  \caption{Integrated flux densities of \deml. The flux density was estimated using images from
    \citetads{1984AuJPh..37..321M} at
    $\lambda$=36~cm and from \citetads{2007MNRAS.382..543H} at $\lambda$=20~cm.}
  \label{tbl-1}
  \begin{tabular}{@{}ccccc}
    \hline
    $\nu$ & $\lambda$ & Beam Size        & R.M.S & S$_\mathrm{Total}$  \\
    (MHz) & (cm)      & (\arcsec)        & (mJy) & (mJy)               \\
    \hline
\phantom{0}843 & 36        & 43.0$\times$43.0 & 0.50  & 154                 \\  
    1377 & 20        & 40.0$\times$40.0 & 0.50  & 146                 \\
    4800 & \phantom{0}6      & 17.3$\times$14.0 & 0.11  & 91                 \\
    8640 & \phantom{0}3      & 17.3$\times$14.0 & 0.06  & 72                 \\
      \hline
  \end{tabular}
\end{table}

\subsection{\ion{H}{I} 21~cm}

To investigate the relative position of the SNR and the superbubble, we
investigated the 21~cm emission line map of neutral hydrogen. We downloaded combined ATCA and Parkes \ion{H}{I} 21~cm radio data
in the direction of \deml\ from the Magellanic Cloud Survey web
page\footnote{\href{http://www.atnf.csiro.au/research/lmc_h1/index.html}{http://www.atnf.csiro.au/research/lmc\_h1/index.html}}. More information about the survey can be found in \citetads{1998ApJ...503..674K,2003ApJS..148..473K}. 
Figure~\ref{FigRadio21cm} shows the neutral hydrogen in an area of
$22\arcmin \times 17\arcmin$ ($\sim$\,320~$\rm pc \times 247~\rm pc$)
with a heliocentric velocity of $\sim$\,264~km/s, which we
  chose since it shows a clear shell-like structure. 
As can be seen in Fig.~\ref{FigRadio21cm}, the position of the SNR and the possible superbubble coincide with an area of
low \ion{H}{I} emission. Around them, a shell-like structure with stronger
\ion{H}{I} emission is visible that surrounds both
objects. This shell can also be seen in Fig.~\ref{FigRadioXY}, which shows the \ion{H}{I} data for
different velocity slices between 233--327~km/s.
The shell is visible at velocities from 258--277~km/s (images in the second row). This suggests the
existence of a hydrogen shell that surrounds both objects,
indicating that both objects are close to each other. 

The \ion{H}{I} catalogue in \citetads{2007ApJS..171..419K} 
assigns a velocity of 276.24~km/s ($\sigma$ = 7.798~km/s) to the
large \ion{H}{I} cloud projected towards \deml, and \citetads{1999ApJ...518..298P} 
determined the heliocentric velocities of the front- and backside of SGS-LMC~2 to be $\sim$\,250~km/s
and $\sim$\,300~km/s, as can be seen in Fig.~\ref{FigRadioXY}, e.g., at 302~km/s to 312~km/s, with \deml\ being projected along the northern rim of this
supergiant shell.

Out of the Magellanic Cloud Survey \hi\ data, we  
created a \hi\ map at a certain position corresponding to a narrow line in the
projected sky over a range of velocities, which we call position-velocity
plots. These plots are presented 
in Fig.~\ref{FigRadioPV} and show the intensity of the \hi\ emission as a
function of the velocity of the gas within the box of Fig.~\ref{FigRadio21cm}
plotted against its RA and its Dec position
(left and right side of Fig.~\ref{FigRadioPV}).
For every position-velocity plot, we combined two pixel rows (or
  columns) of
  Fig.~\ref{FigRadio21cm}. 
According to the resolution of the Magellanic Cloud
  Survey, the velocity information has a resolution of 1.649~km/s. We
  chose the velocity extensions of the boxes that indicate the position of the
  SNR and of the superbubble based upon the
existence of a radio shell at the respective 
velocity slice of Fig.~\ref{FigRadioXY}.
In these position-velocity plots, an enhanced emission can be
  recognized at velocities that \citetads{1999ApJ...518..298P} 
  determined for the borders of SGS-LMC~2, visible as horizontal lines in the plots of
Fig.~\ref{FigRadioPV}a)--d). Furthermore, a circular structure of enhanced emission
is visible in the plots of Fig.~\ref{FigRadioPV}ix)--xv), which corresponds to
the projected Dec position of the YSO \object{2MASS J05425375-6857116} that lies north-west of
\deml\ with respect to the position of the SNR and the superbubble. This
feature is centred at a velocity of $\sim\,275$~km/s and has an extent of
$\pm$8~km/s.

At the position of the SNR and the superbubble, the position-velocity plots show an enhanced 21~cm line emission
feature at the position where a common shell of neutral hydrogen around both
objects would be
expected to lie within projection. 
Although there is evidence for a hydrogen shell, the
data do not allow us to distinguish between a common shell around both objects or two separate
shells. Since in particular the border of the
shell that lies at lower velocities cannot be clearly distinguished, the data
are ambiguous and require further studies, which goes beyond the scope of this paper.


\section{Spectral analysis}
\label{Sect_SpectralAnalysis}

\begin{figure}
  \includegraphics[width=0.5\textwidth]{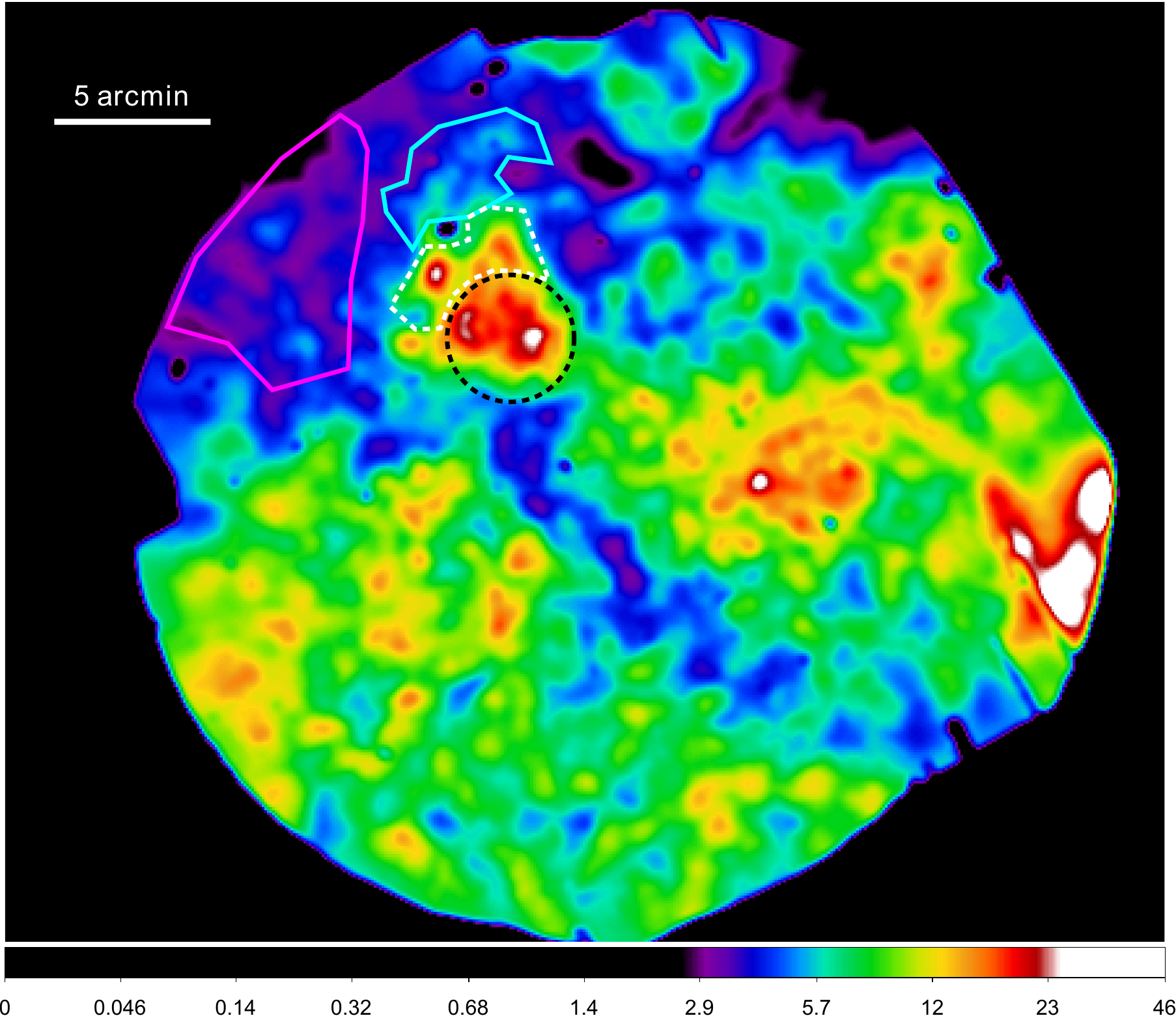}
  \caption{FOV of the \xmm\ observation 0094410101 in an energy band of
    0.8--1.5~keV, smoothed by a Gaussian with a kernel radius of three
    pixels. The colour-scale is in units of
    counts/s/deg$^2$. The marked areas are the fitting regions used for the
    spectral fits of the SNR (black), the part of the superbubble that
    does not overlap the SNR (white), the blowout of the superbubble (cyan),
    and the local background (magenta).}
  \label{FigFittingRegions}
\end{figure}

We analysed the \xmm\ EPIC spectra to investigate the properties of
the hot gas of \deml. The fitting was performed with
Xspec\footnote{\href{http://heasarc.nasa.gov/xanadu/xspec/}{http://heasarc.nasa.gov/xanadu/xspec/}}
version~12.7.1 \citepads{1996ASPC..101...17A}. 
The model we used for the background is based upon the example
given in the ESAS cookbook (see footnote~\ref{FootESAS}).
To estimate the X-ray background, our model includes an
unabsorbed thermal component (Xspec-model: \textit{apec})
for the Local Hot
Bubble with $k_{\mathrm{B}}T=0.1$~keV, two thermal components that are absorbed by the
Milky Way (Xspec-models: \textit{phabs} $\times$ (\textit{apec} + \textit{apec})) for the cooler halo
emission with $k_{\mathrm{B}}T=0.1$~keV and the hotter halo 
with $k_{\mathrm{B}}T=0.5$~keV.
The unresolved extragalactic background was modelled using an absorbed power law
(Xspec-model: \textit{powerlaw}) with
a spectral index $\alpha = 1.46$ \citepads{1997MNRAS.285..449C}, 
which is absorbed by the Milky Way
(Xspec-model: \textit{phabs}) and the LMC (Xspec-model:
\textit{vphabs}) ISM. For the Milky Way, an
$N_{\rm H, MW}$ of $0.06 \times 10^{22}$~cm$^{-2}$ was assumed
\citepads{1990ARA&A..28..215D}. 
The chemical abundances for the LMC were set to $50$~\% of solar abundances
\citepads{1992ApJ...384..508R}. 
This X-ray background was estimated by simultaneously fitting the
used LMC observation and the three combined SEP observations, with frozen
energies, photon index, and $N_{\rm H, MW}$, and with linked normalisations
between all detectors and observations. 

To account for instrumental fluorescence lines, Gaussians with zero
width are included in the
model, e.g. for MOS1 and MOS2
at $1.49$~keV (\element[][]{Al}~K$\alpha$) and $1.75$~keV
(\element[][]{Si}~K$\alpha$), and for pn at $1.49$~keV
(\element[][]{Al}~K$\alpha$). The more energetic fluorescence lines were also
fitted, but since the emission above $\sim$\,4~keV is irrelevant for
these soft X-ray studies, they are not listed here. 
The energies were fitted and then frozen, while their normalisations
remained free for every detector and observation.
Another Gaussian was added at $0.65$~keV to account for possible \ion{O}{VIII}
emission caused by solar wind charge exchange (SWCX) as
reported in \citetads{2009ApJ...697.1214K} 
for all three SEP observations. 

The residual soft proton spectrum was fitted in a separate power-law
model for each detector and each observation. The power
laws are not folded with the instrumental
response matrix, but instead through a diagonal unitary response matrix. While their normalisations were fitted freely, the statistics of
the data were not high enough to do the same for the photon
index. Hence, the
photon index of each power law was set to one, lying within the reasonable
range between 0.2--1.3 (see ESAS cookbook v4.3).

To model the background that was determined by the SEP data, the SEP spectrum
was multiplied by a factor to correct for the different areas.
As a test for the appropriateness of our background estimation through the SEP
observations, we extracted the spectrum of a region as local background within the same pointing as
the superbubble and SNR. This local background region is marked in
magenta in Fig.~\ref{FigFittingRegions} and lies east of the superbubble and
SNR in a region of the FOV as empty as possible. We fitted the spectra of this region simultaneously with the combined SEP
observations to detect deviations. The local background spectrum has a
slight excess, indicating that an additional thermal component is necessary, probably
as a result of local hot ISM in the busy FOV. The fitting results are
listed in Table~\ref{TabFitWerte}. However, the statistics
of this region are far too low to take this area for background estimation instead
of the SEP observations, but we consider its contribution in the
following discussion.

Since the projections of the SNR and the superbubble strongly overlap (cf. Fig.~\ref{FigHardXray}), there should be an
additional emission component of the superbubble in the spectrum of the SNR
and vice versa. To take this into account, we defined a
polygonal region for the fits of the
superbubble that does not overlap the
remnant as defined by the \siihalpha\ data, so that no additional emission
component accounting for the SNR emission is needed in the model for this
polygonal superbubble-only region. We first fitted this superbubble-only spectrum. Since the SNR has well-defined borders through the
circular structure in the \siihalpha\ image, we took this circle as our
extraction region for the SNR (see
Fig.~\ref{FigFittingRegions} for the fitting regions). To investigate a possible blowout of the superbubble to the north-east, as strongly indicated by the optical and X-ray data, we performed a spectral
analysis of this blowout region as marked in cyan in
Fig.~\ref{FigFittingRegions}, which shows the whole FOV of the
\xmm\ observation in an energy range of 0.8--1.5~keV. 

\begin{figure}
  \includegraphics[angle=270,width=0.5\textwidth]{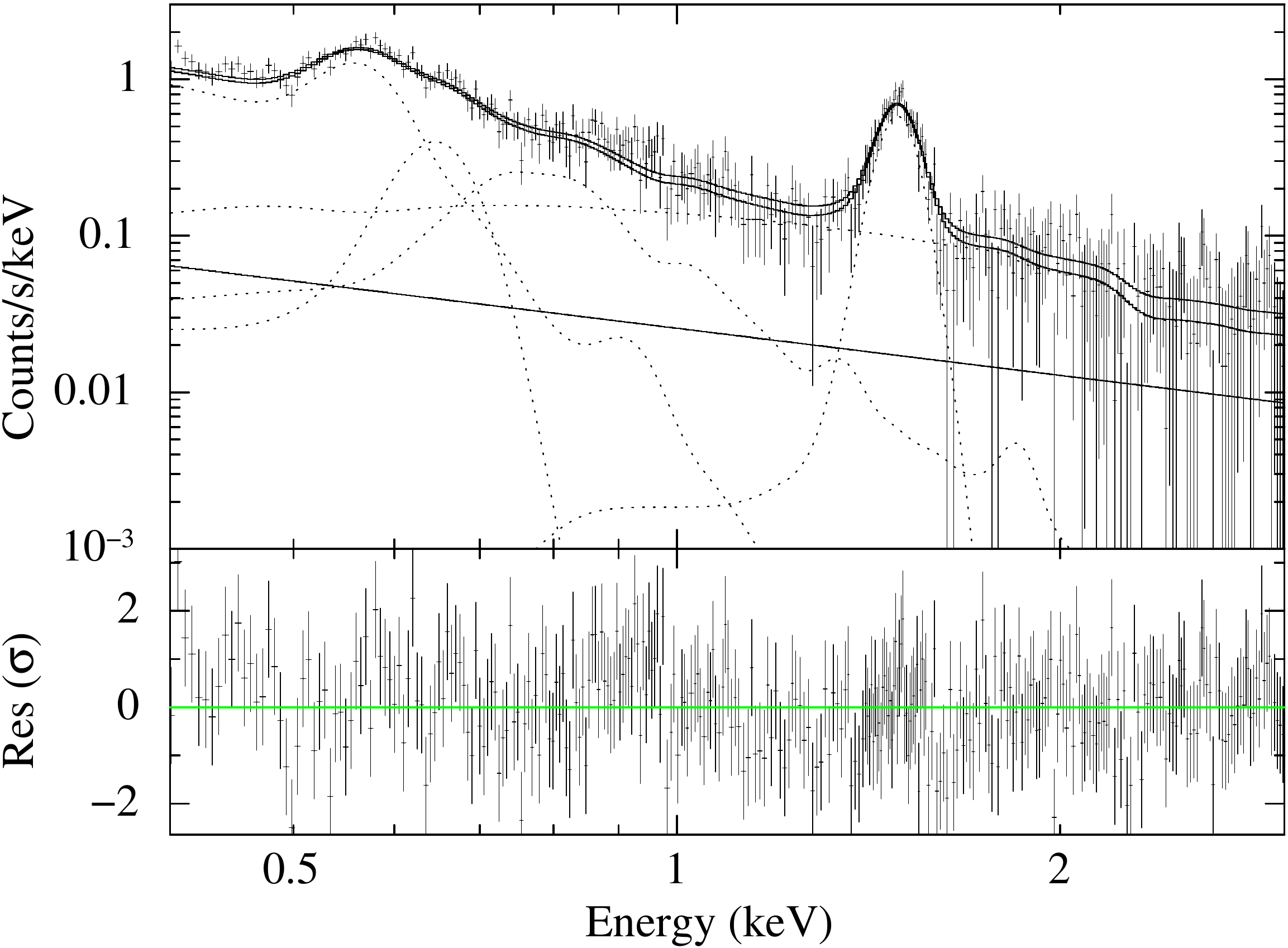}
  \caption{Fitted pn spectrum of the SEP observation used for estimating the
   local X-ray background in an energy
   range of 0.4--3.0~keV with model components and residuals. This spectrum has been 
   fitted simultaneously with the
   SNR spectrum shown in Fig.~\ref{FitSpectraSNR}. 
   The straight line is the modelled residual soft proton
   component. Crosses are data points, while the solid line shows the fitted
   model. The model components are dotted.
}
  \label{FitSpectra}
\end{figure}

\begin{figure*}
  \includegraphics[angle=270,width=0.5\textwidth]{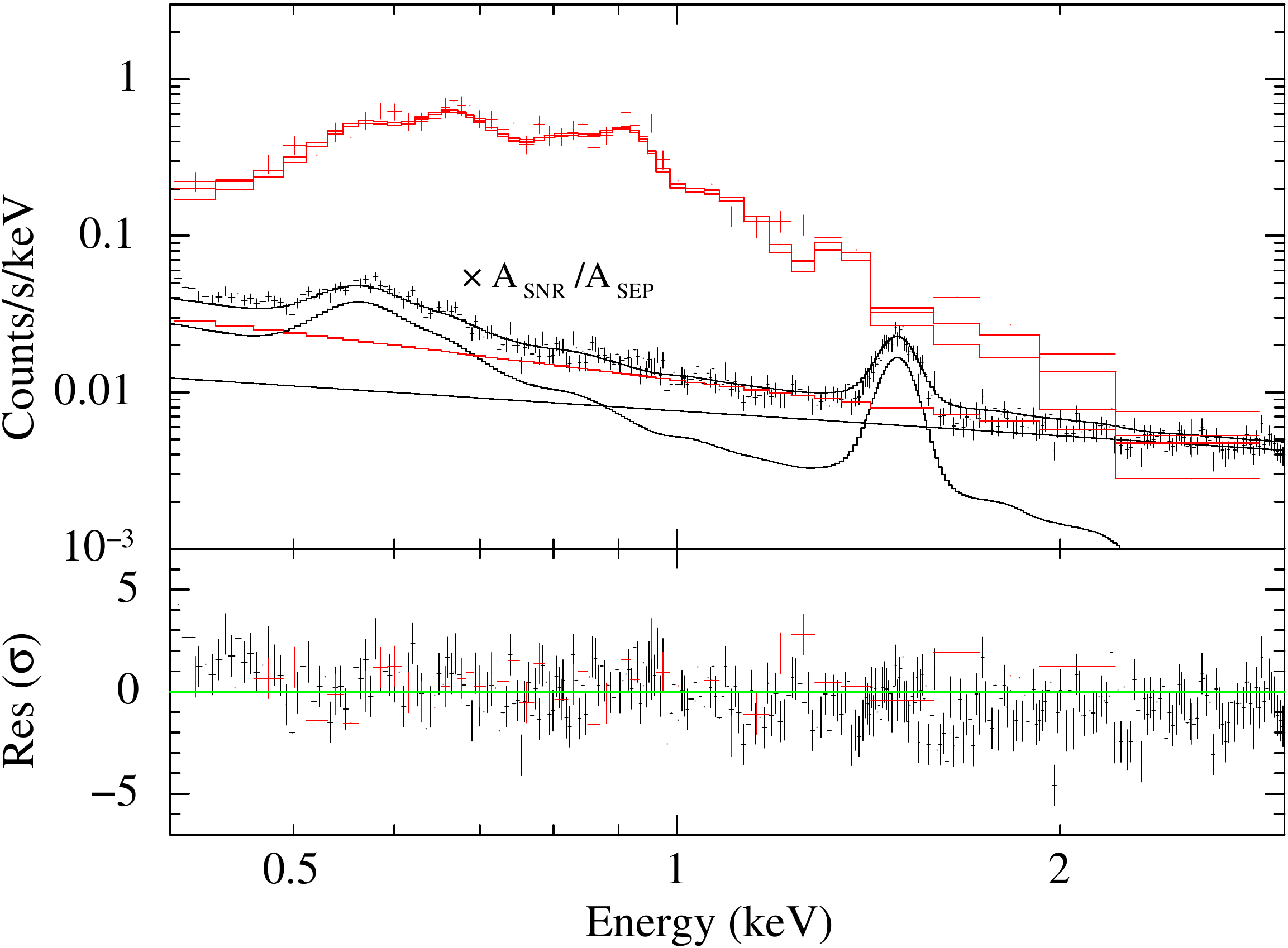}
  \includegraphics[angle=270,width=0.5\textwidth]{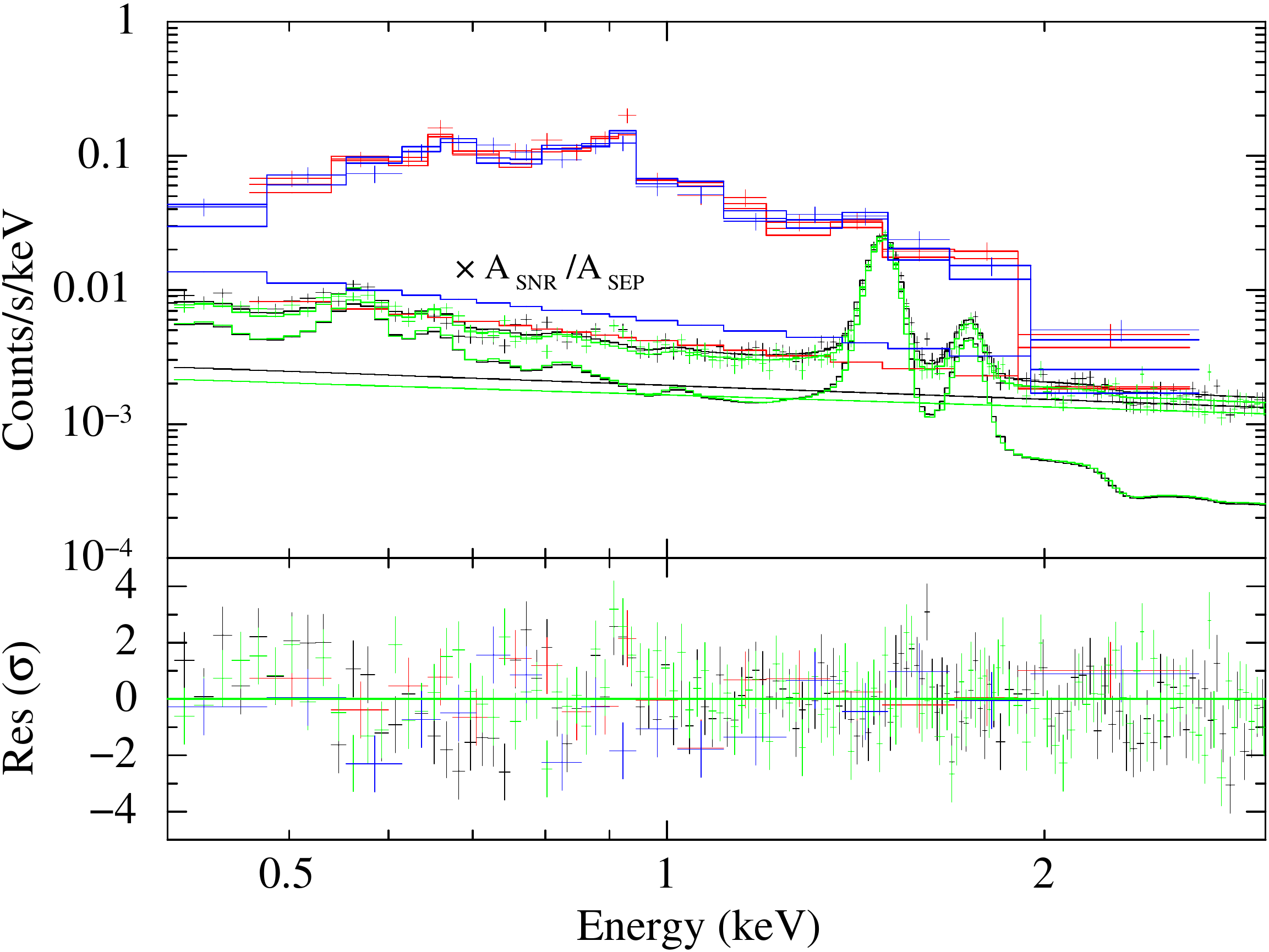}
  \caption{Fitted source and background spectra for the SNR for pn (left) and 
    MOS1/2 (right) with
    residuals in the energy range of 0.4--3.0~keV. The component for the SNR
    emission was fitted with a \textit{vpshock} model. The straight lines are the
    modelled residual soft proton components for the corresponding detector
    and observation. Crosses stand for data points, while the solid lines are
    the fitted models. The source spectra are plotted in red (pn and MOS1)
    or blue (MOS2) and the the background spectra in black (pn and MOS1) or green
    (MOS2). For better presentation, the background spectra are scaled down to the 
    area of the SNR region, which was much smaller than the extraction region for
    the SEP spectrum.
}
  \label{FitSpectraSNR}
\end{figure*}

For the spectra, the data were binned to a minimum of 50~counts per channel and were
fitted in an energy range of 0.3--10.0~keV. We first fitted the combined SEP observations
simultaneously with the individual \deml\ regions to estimate the X-ray photon
background. For the source component, we first tried the \textit{apec} model \citep{2001ApJ...556L..91S}, 
which is a
thermal plasma model with a bremsstrahlung continuum and line emission. However, no good fits were achieved. We then tried two different non-equilibrium models:
a non-equilibrium ionisation plasma model
(Xspec model: \textit{vnei}, \citet{2000RMxAC...9..288B}), 
and a plane-parallel shocked plasma model (Xspec model: \textit{vpshock},
\citet{2000RMxAC...9..288B}). 
For the parameters of these two components, we fixed
the metal abundances at 50~\% of the solar abundances, appropriate
for the LMC. 

To fit the SNR spectrum, we added an additional
thermal emission component to our model to account for the superbubble emission in
this region. We froze the temperature and ionisation timescale $\tau$ of
this superbubble component to the fitted values we derived from our fits of the polygonal
superbubble-only region. The normalisation was frozen as well after multiplying
it with a factor to account for the different sizes of the polygon and the
approximate overlap region of the superbubble with the SNR fitting region.

Since the large number of included model components bear an uncertainty, we performed a series of fits in which we tested more than 20
different temperatures in the range between 0.1~keV and 7~keV for both the
superbubble and the remnant. This resulted in several favoured sets of
fitting results, which we compared based on the goodness of the fit and the ability
to determine the 90~\% uncertainty limits for the parameters. Compared with these two regions, the
SEP data have a much higher count statistic and therefore dominate the goodness of
the fit (cf. Figs.\ref{FitSpectra}, \ref{FitSpectraSNR}). 
To obtain a more meaningful fit of
the superbubble and remnant spectra, we first fitted the entire data as
described before, then froze all background components at these fitted values,
removed the SEP data, and just fitted the \deml\ spectra with the previously
estimated and frozen background components. With this method, we now obtained
more meaningful reduced chi-squared for the different sets of fitting results.

\begin{table*}
 \caption{Resulting best-fit parameters for the polygon (i.e. part of the superbubble), SNR, superbubble
   blowout, and local background region. For all regions, the best-fit values were obtained with the
   \textit{vpshock} model. The column density $n_{\mathrm{H, LMC}}$ is
   given in units of $10^{22}$~cm$^{-2}$, the temperature $k_{\mathrm{B}}T_{\mathrm{x}}$ in
   keV, the ionisation timescale $\tau$ in s~cm$^{-3}$, and the normalization
   $norm$ in cm$^{-5}$. The parameter $norm$ is the normalisation of the
     Xspec model \textit{vpshock} and is a function of the emission measure. In the last row, the reduced $\chi^2$ of the
   fit is given. The uncertainties given are the upper and lower 90~\%
   confidence ranges.}       
  \label{TabFitWerte}      
  \centering                   
  \begin{tabular}{lllll}
     \hline\hline 
     & Polygon                                        & SNR      & Blowout    &   Local Bkg \\
     \hline
     \hbox
         {$nH_{\mathrm{LMC}}$}& 0.30 \tiny(0.07--0.66) & 0.66 \tiny(0.59--0.77)     &
         0.27 \tiny(0--1.2)   & 0.07 \tiny($<$0.3)\\
         $k_{\bb}T_{\mathrm{x}}$   & 0.74 \tiny(0.44--1.1)                               &0.64 \tiny(0.44--1.37) & 0.62 \tiny(0.2--1.0)           & 0.69 \tiny(0.62--0.82)\\
         $\tau$ & 8.5 \phantom{0}\tiny(2.6--28)\,$\times$\,10$^{10}$    & 2.1 \phantom{0}\tiny(1.5--3.4)\,$\times$\,10$^{10}$ & 3.7
         \phantom{0}\tiny(0.25--15)\,$\times$\,10$^{11}$  & 4.6 \phantom{0}\tiny(1.4--11)\,$\times$\,10$^{11}$    \\
         $norm$ & 1.7 \phantom{0}\tiny(1.28--7.5)\,$\times$\,10$^{-4}$     & 8.3 \phantom{0}\tiny(1.8--17.3)\,$\times$\,10$^{-4}$    & 1.2
         \phantom{0}\tiny(0.5--63.1)\,$\times$\,10$^{-4}$    & 1.22 \tiny(0.93--1.6)\,$\times$\,10$^{-4}$ \\
         Red. $\chi^2$ & 0.8                                             & 1.2                   & 0.9          & 1.3                  \\
     \hline                                 
  \end{tabular}
\end{table*}

For the superbubble, we achieved a best-fit result with the \textit{vpshock}-model with
a temperature of $k_{\bb}T_{\mathrm{x, SB}} = (0.74^{+0.36}_{-0.30})$~keV 
and an ionisation timescale of $\tau_{\mathrm{SB}} = (8.5 ^{+19.5}_{-5.9})\times
10^{10}$~s~cm$^{-3}$. 
The errors represent 90~\% confidence intervals 
throughout the paper. For the foreground column density of the LMC
without the Milky Way contribution, we determined a value
of $n_{\mathrm{H, LMC, SB}} = (0.30 ^{+0.36}_{-0.23})
\times 10^{22}$~cm$^{-2}$. 
Table~\ref{TabFitWerte} lists these best-fit
parameters for the superbubble with lower and upper
90~\% limits and reduced chi-squared.

We used this result for the superbubble as an additional component in the
model for the SNR as described above to account for the superbubble contamination in the SNR spectrum. 
With the same method as for the superbubble, our best-fit for the remnant was
again a \textit{vpshock} model and 
resulted in a temperature of $k_{\bb}T_{\mathrm{x, SNR}} = (0.64
^{+0.73}_{-0.20})$~keV, 
an ionisation timescale of $\tau_{\mathrm{SNR}} = (2.1 ^{+1.3}_{-0.6})\times
10^{10}$~s~cm$^{-3}$, 
and an LMC column density of $n_{\mathrm{H, LMC, SNR}} = (0.66 ^{+0.11}_{-0.07})
\times 10^{22}$~cm$^{-2}$ (see Table~\ref{TabFitWerte}). 
The corresponding spectrum of the SNR fit can be seen in 
Fig.~\ref{FitSpectraSNR}. 

For the blowout, we used the same
fitting procedure as described above for the superbubble. We obtained
similar best-fit values for this region as for the superbubble; these are consistent
with a blowout. The blowout is expected to be slightly cooler than the superbubble, but the poor statistics of the observation led to high uncertainties of the resulting
fit and prevented a good measurement of these parameters. The best-fit model was a \textit{vpshock} model with a
temperature of $k_{\bb}T_{\mathrm{x, BL}} = (0.62 ^{+0.38}_{-0.42})$~keV, 
a column density of $n_{\mathrm{H, LMC, BL}} = (0.27 ^{+0.93}_{-0.27})
\times 10^{22}$~cm$^{-2}$, 
and an ionisation timescale of $\tau_{\mathrm{BL}}
= (3.7 ^{+11.3}_{-3.45})\times 10^{11}$~s~cm$^{-3}$ (see Table~\ref{TabFitWerte}).

We calculated the luminosities in the energy band from 0.5--8~keV for the SNR,
superbubble, blowout, and the local background fitting
regions. The resulting
luminosities for the different fitting regions can be found in
Table~\ref{Tab_Lumin}. For the SNR, we obtained an
unabsorbed luminosity of $(1.5 ^{+1.6}_{-1.2}) \times 10^{36}$~erg~s$^{-1}$, 
after subtracting the local background we obtain the same rounded value: $(1.5 ^{+1.6}_{-1.2}) \times
10^{36}$~erg~s$^{-1}$. 
For the superbubble, we added the values for the
polygon and the superbubble contribution in the SNR region, and obtained
$(3.7 ^{+8.5}_{-1.3}) \times 10^{35}$~erg~s$^{-1}$ 
(local background subtracted: $(3.4 ^{+7.8}_{-1.2}) \times 10^{35}$~erg~s$^{-1}$), 
and for the blowout region $(8.3 ^{+425.8}_{-4.9})\times 10^{34}$~erg~s$^{-1}$ 
(local background subtracted: $(5.6 ^{+286.5}_{-3.3}) \times
10^{34}$~erg~s$^{-1}$). 
In the rest of this paper, the unabsorbed luminosities with local
background subtraction are used, if not stated differently. 

\begin{table*}
 \caption{X-ray luminosities in the 0.5--8~keV energy band for the fitting
   region of the polygon (i.e. part of the superbubble), blowout, local background, and SNR. The
   luminosity of the SNR
   fitting region is separated into the contributions of the SNR, the
   superbubble, and the total luminosity of this region. The uncertainties given are the upper and lower 90~\%
   confidence ranges.}       
  \label{Tab_Lumin}      
 \centering                   
  \begin{tabular}{l@{}ccccccc}
     \hline\hline 
     Fitting region          &                            &  Polygon & Blowout &  Local Bkg.&   \multicolumn{3}{c}{SNR}  \\
     \cline{6-8}
                             &                            &          &         &           &   Total &  SNR        & SB  \\
     \hline 
      Abs. $L_\xx$ & 10$^{34}$ [erg~s$^{-1}$]  &  7.9 \tiny(7.7--34.6)    & 3.4 \tiny(1.4--179) & 5.0
      \tiny(3.8--6.6)  & 14.8 \tiny(3.2--30.8)  & 12.9 \tiny(2.8--26.9) & 2.0 \tiny(0.4--4.1)  \\
      Unabs. $L_\xx$ &  10$^{35}$ [erg~s$^{-1}$]         & 2.4 \tiny(1.8--10.7)  &
      0.8 \tiny(0.3--43.4) & 0.8 \tiny(0.5--1.0)    &  16.0
      \tiny(3.4--33.3)
       &  14.7 \tiny(3.1--30.6)   & 1.3 \tiny(0.2--2.7)     \\
      Loc. bkg.\\subtr. $L_\xx$ & 10$^{34}$ [erg~s$^{-1}$]         & 6.5
      \tiny(4.9--22.8) 
      & 1.6 \tiny(0.6--85.7)  &    0      & 12.9 \tiny(2.8--26.9)
       & 11.7 \tiny(2.5--24.9)   & 1.2 \tiny(0.2--2.6) \\
      Unabs. loc.\\bkg. subtr. $L_\xx$ & 10$^{35}$ [erg~s$^{-1}$] & 2.2
      \tiny(1.6--9.8) 
       & 0.6 \tiny(0.2--29.2) &    0      &
      15.7 \tiny(3.4--32.7) 
      & 14.5 \tiny(3.1--30.2)   & 1.2 \tiny(0.2--2.5)  \\
    \hline 
  \end{tabular}
\end{table*}


\section{Discussion}
\label{Sect_Discussion}
\subsection{SNR B0543-68.9}
\label{Subsec_SNR}
To derive more properties of the remnant, we performed calculations based on the Sedov-Taylor-von
Neumann self-similar solution,
which describes the evolution of a remnant in its adiabatic phase
\citepads{1959sdmm.book.....S,1950RSPSA.201..159T,vonNeumann1950}, similar to
what has been done in \citetads{2004ApJ...617..322S} and \citetads{2014MNRAS.439.1110B}. 
We assumed a late adiabatic phase for the remnant since, with a radius of
$\sim$\,30~pc, it is relatively large. Furthermore, 
  we obtained a good chi-squared for the fits with the \textit{vpshock} model
  assuming LMC abundances,
  implying that emission from shocked ISM, and not from the ejecta, dominates
  the remnant.

From the normalisation of the {\it vpshock} or {\it vnei} model, we can determine the emission measure ($EM=\int n_{\ee}n_{\rm{H}}dV$) of the X-ray emitting plasma using the relation
\begin{equation}\label{eq_K}
  K=\frac{10^{-14}}{4 \pi D^2}\int{n_\ee n_\HH}\,dV \,,
\end{equation}
with distance $D$ in cm, electron number density $n_\ee$, and hydrogen number 
density $n_\HH$ in cm$^{-3}$. 
This yields $EM = (2.5 ^{+2.7}_{-2.0}) \times 10^{58}$~cm$^{-3}$.

The preshock H density ($n_{\rm{H},0}$) in front of the blast wave can be
  determined from $EM$. Evaluating the emission integral over the Sedov
  solution using the approximation of \citetads{1975ICRC...11.3566K} 
gives

\begin{equation}
EM=2.07\left(\frac{n_{\ee}}{n_{\rm{H}}}\right)n_{\rm{H},0}^{2}V,
\end{equation}
where V is the volume \citepads[e.g.,][]{1983ApJS...51..115H}. 
We determined the volume of the SNR by estimating the radius of the SNR from
the \siihalpha\ image (see Fig.~\ref{FigOpticalPointsSDivH}) as $\alpha_{\mathrm{SNR}} = 113\arcsec
^{+16\arcsec}_{-8\arcsec}$.
Assuming a distance to the LMC of 50~kpc,
we obtain a radius of $R_{\mathrm{SNR}} = (27.4 ^{+4.1}_{-2.2})$~pc. 
Assuming a spherically symmetric shape of the remnant,
we obtain the volume $V_{\mathrm{SNR}}= (2.53 ^{+1.14}_{-0.62}) \times 10^{60}$~cm$^3$. 
Taking $n_{\ee}/n_{\rm{H}}=1.21$ 
yields $n_{\rm{H},0}=(0.063^{+0.037}_{-0.027})$~cm$^{-3}$. 
The preshock density of nuclei is given as 
$n_{0}\sim1.1 n_{\rm{H},0}$, and it
follows that $n_{0} = (0.069^{+0.041}_{-0.030})\,ff\,$~cm$^{-3}$, with $ff$
being the filling factor of gas in the shell. 
 Hence, the SNR
is expanding into a medium whose density is consistent with an \ion{H}{II} region
\citepads{1977ApJ...218..148M}. 
To calculate this density, a compression factor of four was assumed for the shocked gas \citep{Rankine1870,Hugoniot1887,Hugoniot1889}
and that no energy is lost to cosmic rays.

When a shock runs through a low-density medium like the ISM, the collisional ionisation
equilibrium (CIE) is destroyed. This means that 
the ionisation state of the
ions does not correspond anymore to their temperature. The low value of $2.1
\times 10^{10}$~s~cm$^{-3}$ for
$\tau_{\mathrm{SNR}}$ indicates non-equilibrium ionisation (NEI) for the SNR
\citepads{2010ApJ...718..583S}. This is also supported
  by the fact that the CIE model \textit{apec} yielded much poorer fits than
  the NEI model \textit{vpshock} (see Sect.~\ref{Sect_SpectralAnalysis}).  

To investigate how much the gas deviates from temperature equilibrium,
we used the method of \citetads{1978PASJ...30..489I} 
to test whether temperature equilibration was reestablished, that is, if the
ions and electrons have the same temperature. 
\citeauthor{1978PASJ...30..489I} showed a way to estimate the ratio $f_\TT$ between the X-ray temperature and the
shock temperature: $f_\TT = T_{\mathrm{x, SNR}} / T_\s$, where $f_\TT=1$
corresponds to temperature equilibration, while lower values
indicate its absence. This is done by determining the
intersection point between the curve for Coulomb
equilibration $<T_\ee>/T_\s$ and the curve determined by $T_{\mathrm{x, SNR}}/T_\s =
0.043 T_{\mathrm{x, SNR}} (R_{\mathrm{SNR}} n_\zero)^{-1/2} \nu^{7/5}$, where $\nu$ is a reduced
time variable. Analogously to this, we determined the intersection point $f_\TT$ of these
curves using the previously determined values for $T_{\mathrm{x, SNR}}$, $R_{\mathrm{SNR}}$, and
$n_\zero$. We obtained  $f_\TT = T_{\mathrm{x, SNR}} / T_\s = 
0.37^{+0.59}_{-0.35}$. 
This indicates that temperature equilibration has not
been reestablished. However, one should take this value with caution, as
\citetads{1978PASJ...30..489I} 
used Galactic values for the chemical composition of the ISM, taken
 from \citetads{1973asqu.book.....A}, 
whereas \deml\ is located in the LMC. 
We use this temperature ratio $f_{\rm T} =
  0.37$ for further consideration. 

From our fitted X-ray temperature, we determined the shock velocity,
using the relation
\begin{equation}\label{T_s}
  T_\s = \frac{3\bar{m}}{16k_\bb} \mathrm{v}_\s^2 \,,
\end{equation}
with the Boltzmann constant $k_\bb = 1.38 \times 10^{-16}$~erg~K$^{-1}$ and the mean mass per free particle $\bar{m} = 0.61\,m_\pp$
  for a fully ionised plasma,
  which can be obtained out of the mean mass per nucleus $\bar{m}_\nn=1.4\,m_\pp$. 
If we assume $f_{\rm T} = 0.37$ we derive
$T_\s \approx T_{\mathrm{x, SNR}}/0.37$. Therefore, 
we obtain a shock velocity of $\mathrm{v}_\s= [ (16 k_{\mathrm{B}}T_{\mathrm{x, SNR}}) / (0.4 \times 3 \times 0.61
  m_\pp)] ^ {1/2} = 
(1\,200 ^{+1\,200}_ {-700}) $~km~s$^{-1}$. 

We can now use the Sedov-Taylor-von Neumann similarity
solution to determine the age of the SNR and the initial energy of the
explosion out of these values. We obtain the age $t_\SNR$ of the remnant through the relation
\begin{equation}
  \mathrm{v}_\s = \frac{2 R_{\mathrm{SNR}}}{5t_\SNR} \,.   
\end{equation}
Assuming $f_{\rm T} = 0.37$, we obtain an age of 
$t_\SNR = (2 R_{\mathrm{SNR}}) / (5 \mathrm{v}_\s ) = (8.9 ^{+9.2}_{-5.4} )$~kyr. 

For the initial energy $E_\zero$, the following equation holds:
\begin{equation}
\label{Eq_R_E_t}
  R_{\mathrm{SNR}} = \left( \frac{2.02 E_\zero t_\SNR^2}{\bar{m}_\nn n_\zero} \right)^{1/5} \,.
\end{equation}
This leads to an initial energy of the explosion of 
$E_\zero = (1.4 m_\pp n_\zero R_{\mathrm{SNR}}^5) / (2.02 t_\SNR^2) \approx
  4.3\,(<14.0)\times 10^{51}$~erg, consistent with the 
canonical value of $10^{51}$~erg, when assuming $f_{\rm T} = 0.37$. 
Out of the volume of the SNR and the density of
the ambient medium at the time of the explosion, we can estimate the mass $M_{\mathrm{ISM}}$ of
the ISM that has been swept up by the shock front, assuming a spherically
symmetric remnant and a homogeneous density of the ISM:
\begin{equation}
 M_{\mathrm{ISM}} = V_{\mathrm{SNR}}\,\rho_\zero = \frac{4}{3}\pi
 R_{\mathrm{SNR}}^3 1.4 m_\pp n_\zero = 210\,(^{+150}_{-110})~\rm{M}_{\sun}. 
\end{equation}
This large swept-up mass justifies our earlier assumption of a remnant well into the adiabatic phase.

\subsection{DEM~L299 superbubble}

We used our fitting results of the polygonal region and the equations for an
ideal gas, which we assumed for the gas, to calculate the physical properties of the superbubble. We determined the
number density of hydrogen within this region out of the
normalisation $K$ of the \textit{vpshock} model.
Equation~\ref{eq_K} holds with the only difference of that the densities are now
multiplied by a filling factor $\phi$. This filling
factor accounts for the assumed filling fraction of the superbubble volume with hot gas, which can
be assumed to be $\sim$\,1 for a young interstellar bubble
\citepads{2011A&A...528A.136S}. 
For the volume of the polygonal fitting region, we approximated the polygon
through an ellipsoid with
radii of $a=153\arcsec ^{+2\arcsec}_{-1\arcsec}$ 
=~$(37.1 ^{+1.2}_{-1.1})$~pc 
and $b=65\arcsec ^{+6\arcsec}_{-3\arcsec}$ 
=~$(15 ^{+3}_{-2})$~pc, 
estimated from the \halpha\ image, and the depth
$c=108\arcsec \pm 12\arcsec$ =~$(26.2^{+3.1}_{-3.1})$~pc, 
assumed to be identical to the superbubble
radius. This results in an estimate of the volume of the polygonal fitting region of $V_\poly =
4/3 \pi\,a b c = (1.8 ^{+0.5}_{-0.4}) \times 10^{60}$~cm$^3$,
assuming the LMC distance. Since the size of the polygonal extraction region and the size of
the superbubble as defined through the \halpha\ image differ, we additionally calculated a volume for the latter. We
determined a radius of $R_\SB=108\arcsec ^{+12\arcsec}_{-12\arcsec}$
=~$(26.2
^{+3.1}_{-3.1})$~pc 
for the superbubble through the
\halpha\ image, which leads to a volume of $V_\SB= 4/3 \pi R_\SB^3 =
(2.21 ^{+0.79}_{-0.79})\times 10^{60}$~cm$^3$. 
By using the
same assumptions for the particle density as in
Sect.~\ref{Subsec_SNR} and the normalisation constant of the polygon-fit, Equation~\ref{eq_K} yields the
hydrogen number density of 
\begin{equation}\label{eq_n_H}
n_{\mathrm{H, SB}} =\left( \frac{4 \pi K D^2 }{1.2 \times 10^{-14}
    V_{\mathrm{poly}}\,\phi }\right)^{1/2} 
= (0.048 ^{+0.083}_{-0.009})\,\phi^{-0.5}~\rm{cm}^{-3}  
\end{equation}
for the hot gas inside the superbubble, with $\phi$ being the filling factor.
For the gas pressure inside the superbubble, the ideal gas equation
\begin{equation}
\label{eq_p_ideal_gas}
p_{\mathrm{SB,i}}V_\SB=N k_\bb T_\SB
\end{equation}
holds, with the pressure $p_{\mathrm{SB,i}}$ in g\,cm$^{-1}$\,s$^{-2}$, the volume $V_\SB$ of the
superbubble in cm$^3$, the total number of the particles $N$, the Boltzmann
constant $k_\bb$, and the temperature $T_\SB \approx T_{\mathrm{x, SB}}$ of
the gas in Kelvin.
We can include the total number density $n_\tot$
in Equation~\ref{eq_p_ideal_gas} and express it in terms of the electron and
hydrogen number density as $n_\tot=n_\ee + 1.1 n_\HH$. Since $n_\ee = (1.2 +
0.013\zeta_\LMC)\,n_\HH$, 
Equation~\ref{eq_p_ideal_gas} becomes
\begin{equation}
p_{\mathrm{SB,i}}/k_\bb = 2.31 n_\HH T_{\mathrm{x, SB}}\,
\end{equation}
using our fitted X-ray temperature $T_{\mathrm{x, SB}}$ in Kelvin and the hydrogen density
as determined above. This results in $p_{\mathrm{SB,i}}/k_\bb=(9.5 ^{+17.2}_{-4.3})\times 10^{5}\,\phi^{-0.5}$~cm$^{-3}$~K.

The thermal energy of the gas inside the superbubble can be written as
\begin{equation}
E_{\mathrm{th, SB}}=\frac{3}{2} N k_{\mathrm{B}}T_{\mathrm{x, SB}}\,\phi\,,
\end{equation}
with $N$ being the number of particles. With
Equation~\ref{eq_p_ideal_gas}, the thermal energy becomes $E_{\mathrm{th, SB}} = 3/2 p_{\mathrm{SB,i}} V_\SB\,\phi 
=(4.3 ^{+8.1}_{-2.6})\times 10^{50}\,\phi^{0.5}$~erg. 
The mass $M_{\mathrm{SB, i}}$ of the gas
inside the superbubble can be calculated through the number density, the volume and
the average mass per particle:
\begin{equation}
M_{\mathrm{SB, i}}=n_\tot\,\mu\,m_\HH V_\SB \, ,
\end{equation}
with $\mu=0.61$ being the mean molecular weight of a fully ionised gas. Expressing
$n_\tot$ in terms of the hydrogen density, the mass becomes $M_{\mathrm{SB, i}}=2.31\,n_\HH\,\mu\,m_\HH
V_\SB = (130 ^{+230}_{-60})$~M$_{\sun}$.

We made the same calculations for the blowout region, using an ellipse to
approximate the volume of the fitting region for the blowout and for the
blowout itself. For the fitting region, we chose radii of $a=166\arcsec
^{+2\arcsec}_{-5\arcsec}$ =~($40.2 ^{+1.2}_{-1.7})$~pc, 
$b=84\arcsec ^{+8\arcsec}_{-9\arcsec}$ =~$(20^{+4}_{-4})$~pc, 
and $c=108\arcsec ^{+12\arcsec}_{-12\arcsec}$=~$(26.2 ^{+3.1}_{-3.1})$~pc, 
assuming the LMC distance. $c$ is again assumed to be the same as the radius of the
superbubble. This
resulted in a volume of $V_{\mathrm{BL, fit}}= (2.6^{+0.7}_{-0.7})\times 10^{60}$~cm$^3$ 
for the extraction region of the blowout. With the same method as above, we obtained a hydrogen number
density of $n_\mathrm{H, BL}=(0.034 ^{+0.871}_{-0.011})\,\phi^{-0.5}$~cm$^{-3}$ 
for the hot gas inside the blowout fitting
region. Since the size of the blowout differs from the size of its fitting
region, we additionally determined the X-ray size of the blowout and therefore the
estimated volume of the blowout using our medium energy range X-ray image
(Fig.~\ref{FigFittingRegions}), obtaining $a=157\arcsec ^{+11\arcsec}_{-11\arcsec}$
=~$(38 ^{+3}_{-3})$~pc 
and $b=60\arcsec ^{+11\arcsec}_{-7\arcsec}$ =~$(15 ^{+4}_{-3})$~pc, 
 and assumed again $c=108\arcsec ^{+12\arcsec}_{-12\arcsec}$ =~$(26.2 ^{+3.1}_{-3.1})$~pc. 
This
leads to a volume of the blowout as defined by the medium X-ray image of
$V_{\mathrm{BL}}=(1.8 ^{+0.6}_{-0.5}) \times 10^{60}$~cm$^3$. 
This results in a pressure of $p_{\mathrm{BL,i}}/k_\bb=(0.57 ^{+14.50}_{-0.43})\times 10^{6}\,\phi^{-0.5}$~cm$^{-3}$~K, 
a thermal energy of $E_{\mathrm{th, BL}} = (2.0 ^{+54.1}_{-1.8})\times 10^{50}\,\phi^{0.5}$~erg, 
and a mass of $M_{\mathrm{BL, i}}= (70 ^{+1\,861}_{-36})$~M$_{\sun}$ 
for the hot gas inside the blowout. Although the large
uncertainties of the fits of the X-ray spectrum, resulting from the low
statistics, lead to high uncertainties in
these values, we estimated the total thermal energy of the superbubble plus
blowout region of $E_\thermal = (6.3 ^{+54.8}_{-3.2})\times 10^{50} \phi^{0.5}$~erg 
and the total mass of $M_{\mathrm{i}}= (200
  ^{+1\,876}_{-70})$~M$_{\sun}$. 

We determined the dynamic age of the superbubble using the formula
\begin{equation}
R = \alpha (E_{\mathrm{th, SB}}/ \rho)^{1/5} t_{\mathrm{dyn, SB}}^{2/5} \,,
\end{equation}
\begin{figure*}
  \includegraphics[width=0.5\textwidth]{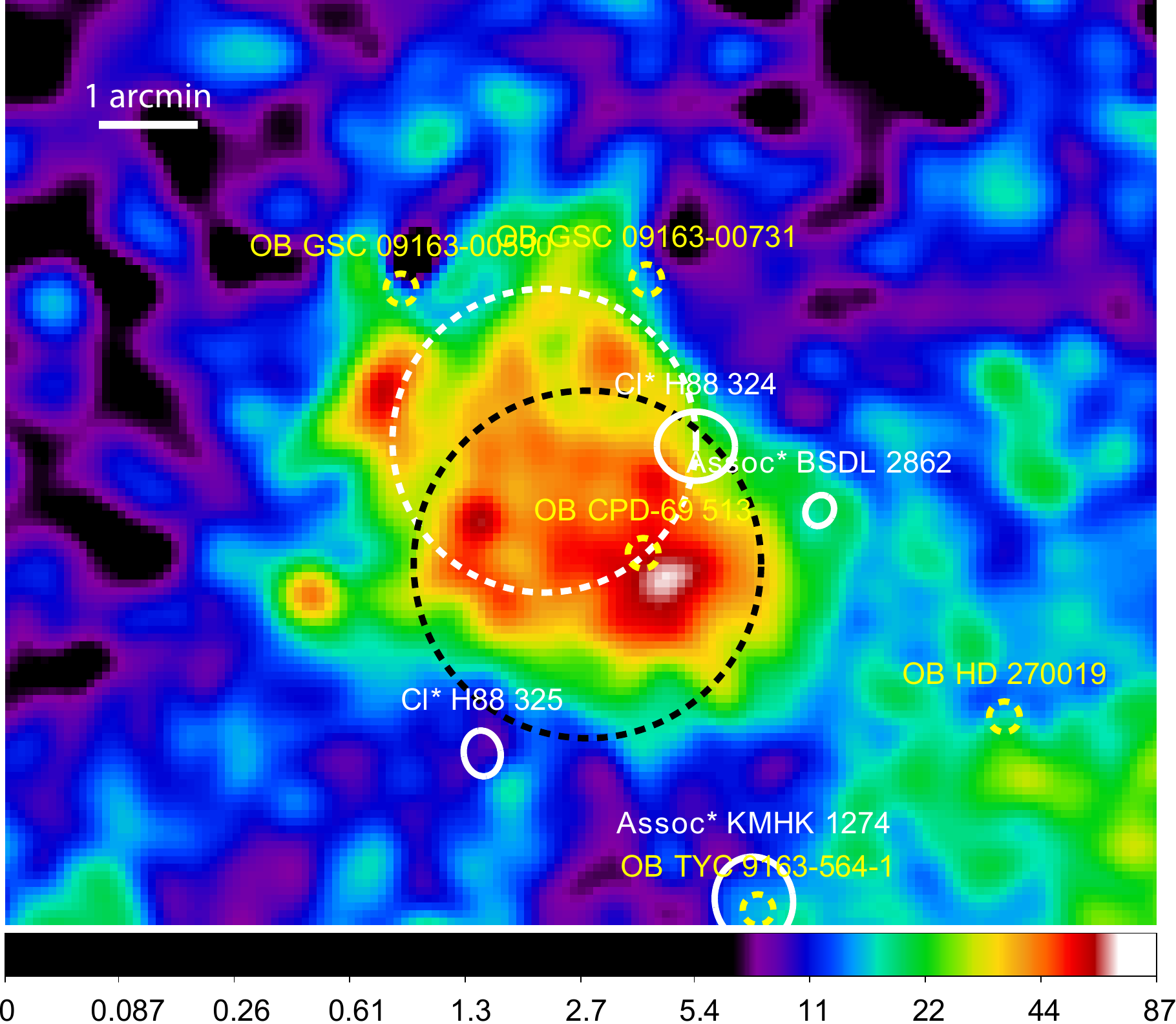}
  \includegraphics[width=0.5\textwidth]{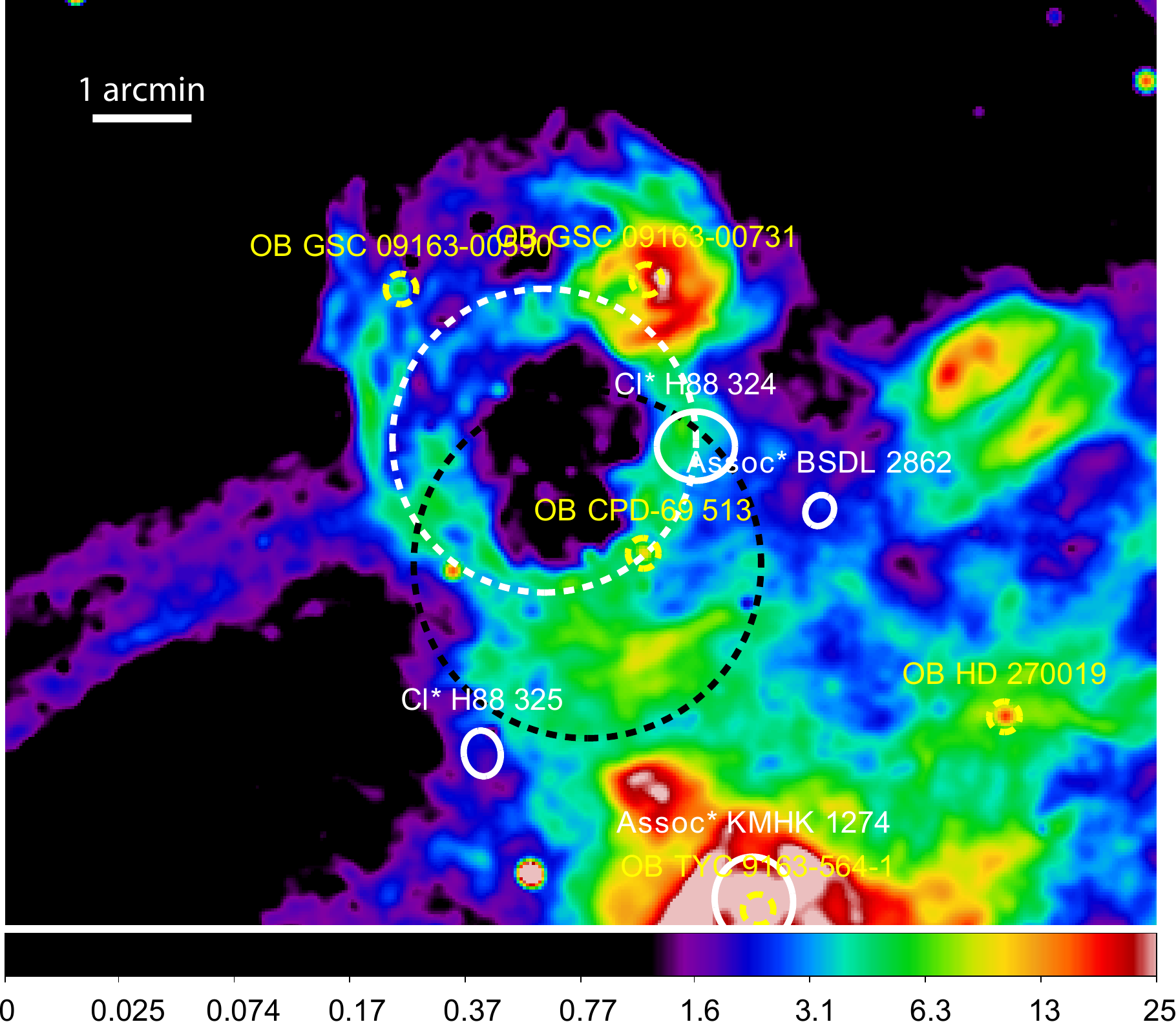}
  \caption{Stellar population in the \deml\ region. Shown are OB-stars
    (yellow dashed small circles), clusters and associations of stars (white
    circles). The position of
    the remnant and of the superbubble are marked with a black and a white
    dashed circle. The images show the region in the broadband (0.3--8~keV)
    X-ray (left) and in the \halpha\ light (right). To create the images, the same
    parameters as in Figs.~\ref{FigAdaptXray}, left side,  and
    \ref{FigOpticalPointsHO}, left side, have been used. Scales
    of the left image are in units of counts/s/deg$^2$, while the scales on
    the right side are in 10$^{-15}$ erg/cm$^2$/s. The population study
    has been performed with SIMBAD.}
  \label{Fig_pop_study}
\end{figure*}
with $R\approx R_\SB$ being the radius of the outer shell in cm, $E_{\mathrm{th, SB}}$ the thermal
energy content of the superbubble in erg, $t_{\mathrm{dyn, SB}}$ the dynamic age of the superbubble in
s, $\rho=n\,\mu\,m_\pp$ is the mass
density of all particles of the ambient medium
in g/cm$^3$, with the particle
density $n$ of the ambient medium, the mean molecular mass $\mu$ and the
proton mass $m_\pp$. This is the same formula as given in Eq.~\ref{Eq_R_E_t},
  but this time with the coefficient $\alpha$ set to 0.76, according
to a bubble in its intermediate stage \citepads{1977ApJ...218..377W}, in
contrast to a coefficient of $2.02^{1/5}\approx1.15$ used for the SNR. 
Solving this equation for $t_{\mathrm{dyn, SB}}$, we derive a dynamic age of
$(70^{+73}_{-35})$~kyr 
for the superbubble, which is younger than the age 
found for other superbubbles. 
Considering the mass and thermal energy of the superbubble as determined above
(without the blowout), this leads to an average mass-loss rate of
$(1.9^{+3.9}_{-1.3})\times 10^{-3}$~M$_{\sun}$/yr 
and an average
energy input rate of $(2.0^{+4.3}_{-1.6})\times 10^{38}$~erg/s. However, this
age seems too young with respect to the stellar ages of the stars that blew
the superbubble. Such young ages for superbubbles have been found in other studies as
well, for example by \citetads{2011A&A...528A.136S} 
or \citetads{2004ApJ...605..751C}. 
This might be explained by the growth rate
discrepancy that has been observed in several superbubbles
\citepads[e.g.][]{2004ApJ...605..751C,2011ApJ...729...28J}. 
It states that, compared with the standard model of
\citetads{1977ApJ...218..377W} 
for stellar bubbles and superbubbles, the growth rate of
the superbubbles is too low regarding the stellar wind
input. 
A discrepancy in the energy budget has been
observed for superbubbles,
since the thermal energy of the hot gas and the kinetic energy of the shells
has been found to be too low to balance the energy input through stellar
winds and supernovae
\citepads{2004ApJ...605..751C,2009ApJ...699..911M}. 
A possible explanation for this
discrepancy is an energy loss of the superbubble through a blowout, as
suggested for this superbubble through optical data. Other possible
explanations are evaporation of cooler, denser cloudlets within the superbubble
\citepads[e.g.][]{2011ApJ...729...28J,1996ApJ...468..722S}. 

\subsection{Input from stars}

\begin{table*}
 \caption{OB-stars, associations of stars, and clusters of stars located within
 a radius of 4\farcm5 from the centre of the SNR as defined by the
 \siihalpha\ image. We have searched SIMBAD for the stellar
   populations in this area. We list the right ascension, declination, identifier and the type of each
 object. The last column indicates a possible physical connection as reported in \citetads{1999AJ....117..238B}.}       
  \label{Tab_pop_study}      
  \centering                   
  \begin{tabular}{lllll}
     \hline\hline 
     RA (J2000)         &   Dec (J2000)         &  Identifier
     &  Type      & Comments \\
     \hline 
     05:42:54.93  &  -68:56:54.5   &  \object{GSC 09163-00731}  &  OB-star   & \\
     05:43:24.67  &  -68:57:00.1   &  \object{GSC 09163-00590}  &  OB-star   & \\     
     05:42:55.414 &  -68:59:52.81  &  \object{CPD-69 513}       &  OB-star   & \\
     05:42:11.561 &  -69:01:39.33  &  \object{HD 270019}        &  OB-star   & \\
     05:42:41.496 &  -69:03:45.10  &  \object{TYC 9163-564-1}   &  OB-star   & \\ 
     05:42:42     &  -69:03.7      &  \object{KMHK 1274}        &  Assoc. of stars
      & in LMC N-164 \\
     05:42:34     &  -68:59.4      &  \object{BSDL 2862}        &  Assoc. of stars
      & \\
     05:43:15     &  -69:02.0      &  \object{H88 325}          &  Cluster of stars & \\
     05:42:49     &  -68:58.7      &  \object{H88 324}          &  Cluster of stars  &
     in \deml\\
     \hline                                 
  \end{tabular}
\end{table*}

We investigated the stellar population in the vicinity of \deml\ to estimate the influence of
the stellar population on the remnant and the superbubble emission. This influence
can be a contribution to the X-ray flux through either OB-stars or
low-mass stars for both objects, as well as energy input and mass input through supernova explosions and stellar
winds for the superbubble.

For this purpose, a population study was performed with the SIMBAD
astronomical database\footnote{SIMBAD:
  \href{http://simbad.u-strasbg.fr/simbad/}{http://simbad.u-strasbg.fr/simbad/}}. Within
a radius of 4\farcm5 (65~pc) from the centre of the remnant as defined by the
\siihalpha\ image, we found five OB-stars, two clusters of stars, and two
associations. These sources can be found in Table~\ref{Tab_pop_study} and are
plotted in Fig.~\ref{Fig_pop_study}, with the position and size of the
clusters and associations as given in the catalogue of
\citetads{1999AJ....117..238B}. 
The association KMHK 1274, which is marked as possibly connected to LHA 120-N 164
by \citetads{1999AJ....117..238B}, and the two OB-stars in the south-west of \deml\ lie too far from the remnant (4\arcmin) and the superbubble (2\arcmin\ and
3\arcmin) to be of much influence. One OB-star is
projected towards \deml\ at the north-western rim of the superbubble close
to a star-forming region. The
remaining two OB-stars, two clusters, and one association lie close to the
borders of the remnant and/or the superbubble in projection, with one OB-star being inside the remnant. The cluster H88 324 was found to be possibly connected
to \deml\ by \citetads{1999AJ....117..238B}. Unfortunately, no deeper study of this cluster or for the other cluster and
association lying close to the superbubble could be found, and no
further information about the spectral types of the OB-stars is listed
in SIMBAD.

\subsubsection{Energy and mass input for the superbubble}
\begin{figure}[h]
 \includegraphics[width=0.5\textwidth]{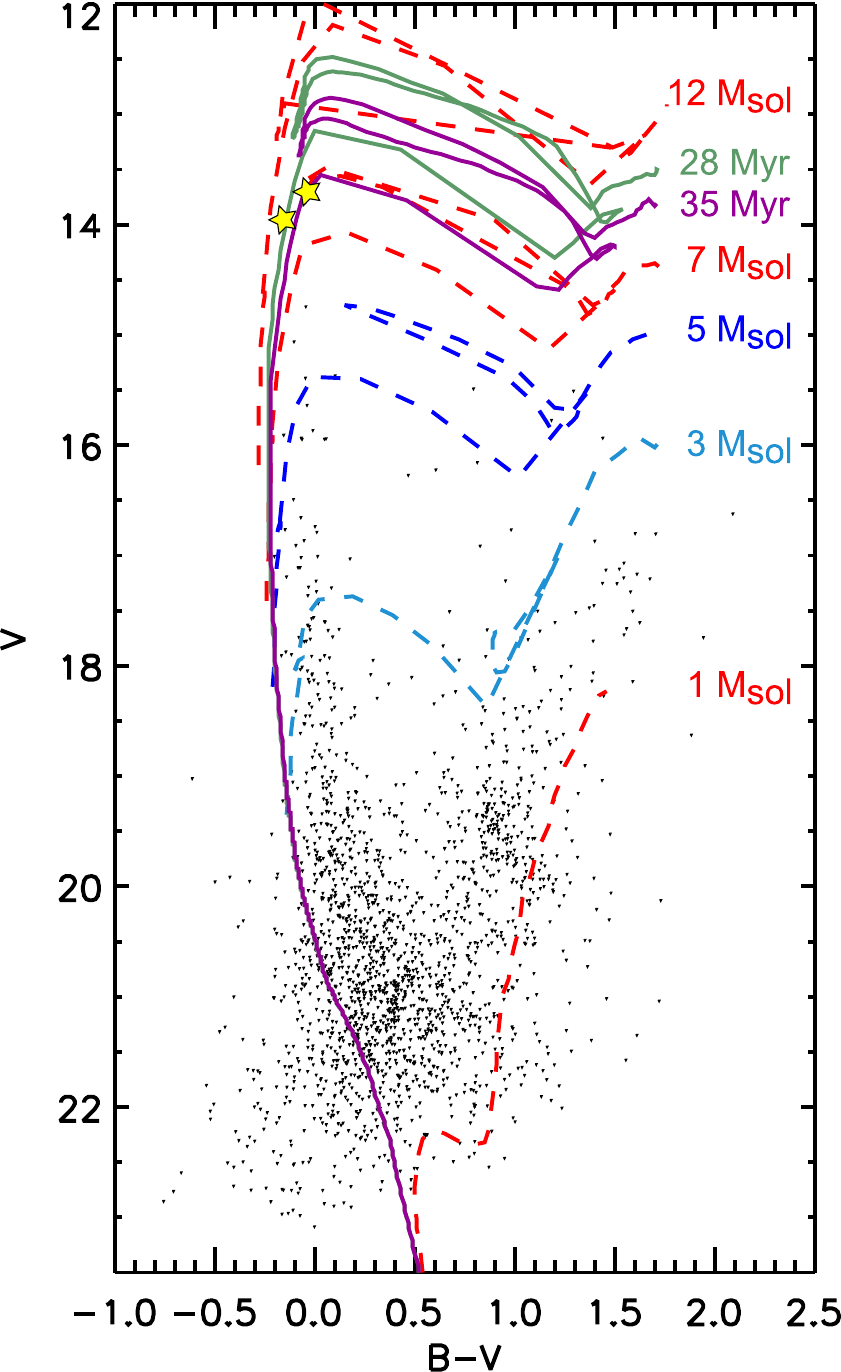}
  \caption{Colour-magnitude diagram for stars projected 2\arcmin\
    around the centre of the superbubble. Plotted are $\sim$\,1\,900 stars from the
    photometric catalogue of
    \citetads{2004AJ....128.1606Z}. 
    The dashed lines show the evolutionary tracks for stars with 1, 3 (light blue), 5 (dark blue), 7 and 12~M$_{\sun}$ (marked in red if not stated
    differently, from bottom to top). The two most massive stars are marked with
    yellow star symbols. Two isochrones at
    28~Myr (green) and 35~Myr (purple), which are taken from
    \citetads{2001A&A...366..538L}, are plotted as solid curves. 
  }
  \label{Fig_CM-Diagram}
\end{figure}
We estimated the energy and mass input of previous supernova
explosions and stellar winds of massive stars to compare them with observed values. For this
estimate, we needed the number of massive stars that were initially formed in
this region, as well as the age of the stellar population.  
We used two different ways to
estimate these numbers: via a colour-magnitude diagram, and via the software
package
Starburst99\footnote{\href{http://www.stsci.edu/science/starburst99/}{http://www.stsci.edu/science/starburst99/}}
\citep{1999ApJS..123....3L,2005ApJ...621..695V}. 
This package allows model predictions for a variety of photometrical and
spectral properties of a population of stars.

For the first method, we created a colour-magnitude diagram using the photometric catalogue of \citetads{2004AJ....128.1606Z} for stars
within a radius of 2\arcmin\ (30~pc) around the centre of the superbubble. 
Figure~\ref{Fig_CM-Diagram} shows the resulting colour-magnitude diagram, including two
isochrones at 28~Myr and 35~Myr (solid lines) as well as the evolutionary tracks for stars with an initial mass of
1, 3, 5, 7, and 12~M$_{\sun}$ from \citetads{2001A&A...366..538L}. 
For these tracks, we chose the basic model set for the Johnson-Cousins-Glass
photometry, calculated for a metallicity of Z=0.008, since this lies close
to the metallicity of the LMC, which is about half of the solar value $Z_{\sun}$=0.02
(see \citetads{2001A&A...366..538L} for more details about the evolutionary
tracks and isochrones). In the diagram, $\sim$\,1\,900
stars are plotted, with the most massive main-sequence stars having approximately
10~M$_{\sun}$. Stars that are obviously separated
from the main sequence but not at the turn-off point from the main
sequence of the stellar population are considered as fore- and background stars and are
therefore not taken into account.

\begin{figure}
  \includegraphics[angle=270,width=0.5\textwidth]{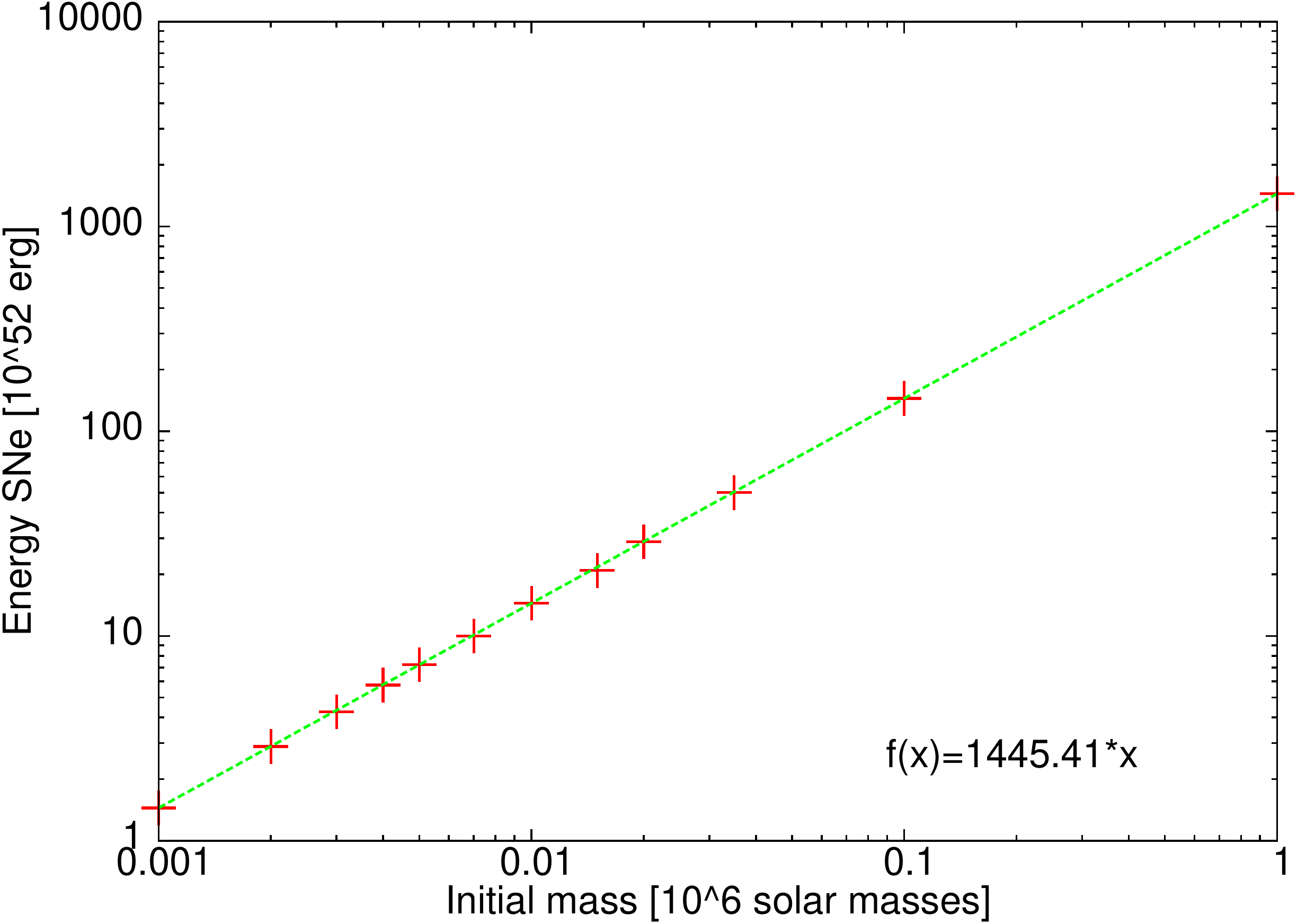}
 \caption{Logarithm of the energy input through supernovae as a function of
   the initial mass of the simulated stellar population with an age of
   32 Myr. The simulation was made using Starburst99. Crosses are the simulation results and the
   dashed line is a fit to these values. This fitted function f(x) has been used to
   extrapolate to masses below the lower initial mass limit.  
 }
  \label{Fig_statischtische_Aussage_ST99}
\end{figure}

Assuming an IMF with a Salpeter power-law exponent of
$\Gamma$=-1.35 \citepads{1955ApJ...121..161S} 
in the mass range from 1--100~M$_{\sun}$, we calculated the number of stars in the mass bin from 8--100~M$_{\sun}$, with 8~M$_{\sun}$ being the lower initial
mass limit for a star to become a supernova. This was based on
numbers in six different mass bins between 3--7~M$_{\sun}$. 
We received an average number of
9$^{+3}_{-3}$ stars for the mass bin from 8--100~M$_{\sun}$. Since two stars lie above the evolutionary tracks for
8~M$_{\sun}$ in the colour-magnitude diagram, we conclude that 7$^{+3}_{-3}$
supernovae should have already occurred in the superbubble region.

We used the colour-magnitude diagram to determine the approximate age of the
population. Since the two most massive stars, which lie between
7--12~M$_{\sun}$ (marked with yellow star symbols),
are both on or close to the main sequence, no turn-off point from the main
sequence could be determined, but an upper age limit of 28--35~Myr can
be inferred for the superbubble population, as indicated by the two plotted
isochrones in Fig.~\ref{Fig_CM-Diagram}. In addition, we determined the total
initial mass for the stars between 1--100~M$_{\sun}$ and obtained an average
total mass of $(636 ^{+157}_{-158})~$M$_{\sun}$. 

As a second way to determine the number of already occurred SNe, we used the
Starburst99 v6.0.4 software
and data package. 
As input parameters, we used the age and the
total mass as obtained from the colour-magnitude diagram, a Salpeter
IMF, and a metallicity of Z=0.008. Since Starburst99 was designed for simulating the population of whole galaxies, it has a lower initial mass
limit of $1 \times 10^3$~M$_{\sun}$ and considers only masses in
steps of $1\,000$~M$_{\sun}$. Since our average initial mass of
$\sim$\,$640~$M$_{\sun}$ lies below the limit of Starburst99, we ran 12 different simulations with
initial masses between $1 \times 10^3$~M$_{\sun}$ and $1 \times
10^6$~M$_{\sun}$ and extrapolated to our lower mass. We found that both the energy input and the total mass loss of the stars are a linear
function of the mass (see Fig.~\ref{Fig_statischtische_Aussage_ST99} for the
energy input of supernovae). We extrapolated an energy input through core-collapse
supernovae of $\sim$\,$(9.1 ^{+3.1}_{-3.1}) \times 10^{51}$~erg 
(errors with respect to the upper and
lower mass and time limits), 
leading to $\sim$\,$ 9^{+3}_{-3}$ 
previous supernovae, assuming an energy input of
$10^{51}$~erg per supernova. This large number of SNe and therefore a large
energy input might be a result of uncertainties in the colour-magnitude
diagram concerning the confusion with fore- and background stars and the error
propagation of this to the input parameters for Starburst99. As a wind input, we derive $\sim$\,$(1.6 ^{+0.4}_{-0.4}) \times
10^{51}$~erg. 
For the total mass loss through supernovae and stellar winds, Starburst99
yields a value of approximately $(164 ^{+48}_{-46})~$M$_{\sun}$. 
This is similar to the mass that we found for the hot gas inside the
superbubble, although the superbubble mass has a very high upper limit. Indeed, the superbubble mass is expected to be higher than the mass loss through stellar winds and supernovae, since evaporation of the swept-up shell \citepads{1996ApJ...468..722S} 
is an important mass contribution to the superbubble.

\subsubsection{Stellar contribution to the X-ray luminosity}
To estimate the X-ray luminosity of the superbubble and the SNR, we have to
determine the amount of X-ray emission originating from stars in these
regions that were not masked as point sources in our previous analysis. Within the
fitting regions for the superbubble and the SNR, there are two catalogued
OB-stars, one in each region. Since an O-star has an X-ray luminosity
of typically $\sim$\,$10^{31}$--$10^{33}$~erg~s$^{-1}$
\citepads{1989ApJ...341..427C,1997A&A...322..167B,2006MNRAS.372..661S}, the high-mass stars account for
$\sim$\,$2 \times 10^{31}$--$2 \times 10^{33}$~erg~s$^{-1}$ of the X-ray luminosity of our X-ray spectra. 
The other stellar contribution of X-ray luminosity that has to be taken into
account are low-mass stars between 0.008--3~M$_{\sun}$. Although the X-ray
luminosity of each of these low-mass stars is lower than that of an OB-star,
these low-mass stars are so abundant that their X-ray emission might be
observed as a diffuse emission \citepads{2005MNRAS.361..679O}.
Therefore, we have to know
their number and average X-ray luminosity per mass bin to be able
to estimate their fraction of diffuse emission in our
X-ray images. To do so, we created a second
colour-magnitude diagram for the area enclosed by a circle with a radius of
2\arcmin\ around the centre of the superbubble and a circle around the SNR with
a radius of 2\farcm3. We obtained a value of $306^{+58}_{-58}$ low-mass stars. 
An estimate of the luminosity per star in a certain mass bin can be found by comparing our
population with the population of the well-studied Orion nebula cluster (ONC),
assuming that both populations differ only in their size and age and have the same
IMF (for this method, see e.g. \citetads{2006ApJS..163..306G,2006ApJ...638..860E,2007ApJS..169..353B,2012A&A...547A..19K}). 
The Orion nebula cluster has been studied in great detail through the \chandra\
Orion Ultradeep Project \citepads[COUP,][]{2005ApJS..160..319G}. 
Using this COUP observation, \citetads{2005ApJS..160..379F} 
determined the number of stars and the integrated luminosity per mass bin for
the ONC as well as the relative contribution of each mass bin
to the X-ray luminosity. For example, in the mass bin from 1--3~M$_{\sun}$,
they reported an integrated, unabsorbed luminosity in the energy band
from 0.5--8~keV of $9.12 \times 10^{32}$~erg~s$^{-1}$, with 70 stars lying in this mass bin. Rescaling these values to the size of our stellar
population, we used our upper limit of 364 stars in the 1--3~M$_{\sun}$ bin
and obtained an integrated luminosity of $4.7 \times 10^{33}$~erg~s$^{-1}$. 
According to \citetads{2005ApJS..160..379F}, 
the X-ray luminosity in this bin contributes 41~\% of the total X-ray
luminosity, while stars below 1~M$_{\sun}$ account for another 33~\%. Using
these numbers, we obtain a total X-ray luminosity of $8.6 \times 10^{33}$~erg~s$^{-1}$ 
for our population of stars below
3~M$_{\sun}$. Out of the colour-magnitude diagram, we
determined an age of 28--35~Myr for the superbubble plus SNR population, which
therefore seems to be older than the ONC population, for
which \citetads{2005ApJS..160..390P} 
derived age estimates of 
$\lesssim$\,10~Myr for their star sample
within the ONC, and \citetads{2005ApJS..160..319G} 
stated that 80~\% of the stars within 1~pc from the centre of the
ONC are younger than 1~Myr. Since the X-ray luminosity of stars below
3~M$_{\sun}$ decreases with time \citepads[see][]{2005ApJS..160..390P}, 
the determined value for the X-ray luminosity of the low-mass stars can be
considered as an
upper limit. Compared with the X-ray luminosity of
$1.79 \times 10^{36}$~erg~s$^{-1}$ 
in the same energy band from 0.5--8~keV that we obtained in Sect.~\ref{Sect_SpectralAnalysis} from our X-ray fits for the
superbubble plus the SNR (i.e. for the polygonal superbubble-only and the SNR fitting-region),
the X-ray luminosity of low-mass stars only accounts for $\sim$\,0.5~\% 
of the diffuse X-ray emission in this region and can therefore be neglected in
the context of this analysis.


\section{Summary}
\label{Sect_Summary}
 
We presented a multi-wavelength study of
\deml\ in the LMC. The morphological study of X-ray, optical, and radio data
revealed that in addition to the supernova remnant SNR B0543-68.9, there is evidence for a superbubble. The position of
the SNR (centre: RA 05:43:02.2, Dec -69:00:00.0 (J2000), radius:
$\sim$\,30~pc) and the superbubble (centre: RA 05:43:07.4, Dec \mbox{-68:58:39.3} (J2000), radius:
$\sim$\,25~pc) were identified using the
[\ion{S}{II}]/H$\alpha$ flux-ratio image of the remnant and through a shell-like
structure visible in the optical data for the superbubble. The two objects show
diffuse X-ray emission, and a cold hydrogen shell-like structure is
visible around both objects in \ion{H}{I} data. The projection of
the objects overlaps, with the superbubble lying slightly farther to
the north. An indication for a blowout of the superbubble at its northern rim was found in
the optical and X-ray data. We found an indication for a common hydrogen
  shell around the two objects, which means that they lie in the vicinity of each
  other, although the data are ambiguous. But since the
\deml\ region is located next to an active star formation region in the LMC
and since the centre of the remnant is projected inside the superbubble, it
seems likely that the two objects are connected and that the progenitor star of
the SNR was part of the stellar association that formed the superbubble. 
Although there is evidence for a hydrogen shell, the data do
not allow us to distinguish whether this is a common shell or two separate shells. Out of the
radio continuum data, we found a spectral index of $\alpha=-0.34$ for the radio emission of SNR
B0543-68.9. Since this index is rather flat, it indicates a dominance of the
thermal emission and therefore a rather mature remnant, which was also
found from the results of the X-ray spectral analysis.

In the X-ray spectral analysis, we fitted three different regions of \deml: a
superbubble-only region, an SNR region that also contains superbubble emission,
which we estimated through the fitting results of the superbubble-only region,
and the blowout region north of the superbubble. For all three regions, the
plane-parallel shock model \textit{vpshock} gave the best fit
results. We obtained a temperature $k_{\mathrm{B}}T$ for SNR B0543-68.9 of
  $(0.64^{+0.73}_{-0.20})$~keV, for the superbubble of
  $(0.74^{+0.36}_{-0.30})$~keV, and for the blowout of
  $(0.62^{+0.38}_{-0.42})$~keV, and determined the luminosities of the
  objects.
Applying the Sedov
solution, we used these results for the remnant to calculate its other
properties, such as an age of $8.9 ^{+9.2}_{-5.4}$~kyr, 
a shock velocity of $(1\,200 ^{+1\,200}_ {-700}) $~km~s$^{-1}$, a swept-up mass of $210\,(^{+150}_{-110})~\rm{M}_{\sun}$, and an initial energy of the
explosion of $4.3\,(<14.0)\times 10^{51}$~erg. For the superbubble, we used the ideal gas equation to determine
properties such as a pressure of $(9.5 ^{+17.2}_{-4.3})\times
10^{5}\,\phi^{-0.5}$~cm$^{-3}$~K, a mass of the hot gas inside the
  superbubble of $(130 ^{+230}_{-60})$~M$_{\sun}$,
a thermal energy content of $(4.3 ^{+8.1}_{-2.6})\times
10^{50}\,\phi^{0.5}$~erg of the hot gas, and a dynamic age of $(70^{+73}_{-35})$~kyr. 

We estimated the influence from stars on our results by estimating their
energy and mass input for the superbubble.
Using the colour-magnitude
   diagram, we obtained a number of $(7^{+3}_{-3})$ past supernovae in the
  \deml\ region and an age estimate of $\lesssim$28--35~Myr for the population
  that created the superbubble. We used Starburst99 to
  extrapolate the mass input through supernovae and stellar winds of $(164 ^{+48}_{-46})~$M$_{\sun}$, and an
  energy input of $\sim$\,$(1.6 ^{+0.4}_{-0.4}) \times
10^{51}$~erg through stellar winds and of $\sim$\,$(9.1 ^{+3.1}_{-3.1}) \times
10^{51}$~erg through supernovae. Thus, considering
  both results, we
  determined the number of already occurred supernovae in the \deml\ region to be 4--12.
Furthermore, we showed that the stellar X-ray
contribution is negligible when estimating the diffuse X-ray
emission of this region, since we found the high-mass and low-mass stars to account
for only 0.6~\% and 0.5~\% of the diffuse emission.


\begin{acknowledgements}
  We thank the anonymous referee for the helpful comments.
    This research has been funded by the Deutsche Forschungsgemeinschaft
    through the Emmy Noether Research Grant SA 2131/1-1.
    P.J.K. acknowledges support by the Deutsche Forschungsgemeinschaft
    through the BMWi/DLR grant FKZ 50 OR 1209. We thank J\"org
    Bayer for his help in optimising Fig.~\ref{FigHalphaBig}.
    This work is based on observations obtained with XMM-Newton, an ESA science mission with instruments and contributions directly funded by ESA Member States and NASA.
    The MCELS project has been supported in part by NSF grants AST-9540747 and AST-0307613, and through the generous support of the Dean B. McLaughlin Fund at the University of Michigan, a bequest from the family of Dr. Dean B. McLaughlin in memory of his lasting impact on Astronomy. The National Optical Astronomy Observatory is operated by the Association of Universities for Research in Astronomy Inc. (AURA), under a cooperative agreement with the National Science Foundation.
    This research has made use of the SIMBAD database, operated at CDS, Strasbourg, France.
 \end{acknowledgements}

\bibliographystyle{aa} 
\bibliography{Bib}


\end{document}